\documentclass[12pt, english]{article}
\usepackage[T1]{fontenc}
\usepackage[latin9]{inputenc}
\usepackage[margin=1in,bottom=1in,top=1in]{geometry}
\usepackage{lscape}
\usepackage{setspace}
\usepackage{amsmath}
\usepackage{amsthm}
\usepackage{mathabx}
\usepackage{bm}
   
\usepackage{amssymb}
\usepackage{stmaryrd}
\usepackage{graphicx}
\usepackage{float}
\usepackage{setspace}
\usepackage{rotating}
\usepackage{subcaption}
\usepackage{ragged2e}
\usepackage{esint}
\usepackage{ulem}
\usepackage{babel}
\usepackage{longtable}
\usepackage{booktabs}
\usepackage{caption,booktabs,array}

\usepackage[round]{natbib}
\usepackage[hypertexnames=false]{hyperref}
\usepackage{xcolor}
\usepackage{colortbl}
\definecolor{darkblue}{rgb}{0,0,.6}
\hypersetup{
colorlinks=true,
linkcolor=darkblue,
filecolor=darkblue,
urlcolor=darkblue,
citecolor=darkblue,
}
\usepackage[toc,page]{appendix}
\usepackage{booktabs}

\usepackage[sc]{mathpazo}
\usepackage{colortbl}
\newcommand{\indep}{\perp \! \! \! \perp}

\newtheorem{assumption}{Assumption}

\newtheorem{remark}{Remark}
\newtheorem{example}{Example}
\newtheorem{theorem}{Theorem}
\newtheorem{lemma}{Lemma}
\newtheorem{proposition}{Proposition}

\newtheorem{corollary}{Corollary}

\usepackage{dsfont}
\usepackage{pifont}

\usepackage{tikz}
\usetikzlibrary{arrows}
\usetikzlibrary{patterns}
  \pgfdeclarepatternformonly{north east lines wide}%
        {\pgfqpoint{-1pt}{-1pt}}%
        {\pgfqpoint{10pt}{10pt}}%
        {\pgfqpoint{9pt}{9pt}}%
        {
            \pgfsetlinewidth{0.3pt}
            \pgfpathmoveto{\pgfqpoint{0pt}{0pt}}
            \pgfpathlineto{\pgfqpoint{9.0pt}{9.0pt}}
            \pgfusepath{stroke}
        }
        \pgfdeclarepatternformonly{north west lines wide}%
        {\pgfqpoint{-1pt}{1pt}}%
        {\pgfqpoint{10pt}{-10pt}}%
        {\pgfqpoint{9pt}{-9pt}}%
        {
            \pgfsetlinewidth{0.3pt}
            \pgfpathmoveto{\pgfqpoint{0pt}{0pt}}
            \pgfpathlineto{\pgfqpoint{9.0pt}{-9.0pt}}
            \pgfusepath{stroke}
        }
\definecolor{green}{rgb}{0.0,0.7,0.0}

\begin{document}
\normalem

\title{Endogenous Heteroskedasticity in Linear Models\thanks{%
We thank Hidehiko Ichimura, Augusto Nieto-Barthaburu, and seminar participants 
at Universidad de San Andr\'es, Universidad de Buenos Aires, and RedNIE for 
very helpful and constructive comments. Computer programs to replicate the 
numerical analyses are available from the authors. All remaining errors are 
our own.

\textit{Alejo:} IECON--Universidad de la Rep\'ublica, Montevideo, Uruguay. 
E-mail: javier.alejo@fcea.edu.uy; 
\textit{Galvao:} Department of Economics, Michigan State University. 
E-mail: agalvao@msu.edu;
\textit{Martinez-Iriarte:} Department of Economics, UC Santa Cruz. 
E-mail: jmart425@ucsc.edu; 
\textit{Montes-Rojas:} Universidad de Buenos Aires, Facultad de Ciencias 
Econ\'omicas, Instituto Interdisciplinario de Econom\'ia Pol\'itica, CONICET, 
and Universidad Argentina de la Empresa (UADE), Instituto de Econom\'ia 
(INECO), Buenos Aires, Argentina. 
E-mail: gabriel.montes@economicas.uba.ar.
}}

\author{Javier Alejo,
Antonio F. Galvao,\\
Julian Martinez-Iriarte, and
Gabriel Montes-Rojas}

\date{\today}

\maketitle
\thispagestyle{empty}
\vspace{-2em}

\begin{abstract}\singlespacing

\noindent Linear regressions with endogeneity are widely used to estimate causal effects. This paper studies a framework that involves two common practical issues: endogeneity of the regressors and heteroskedasticity that depends on endogenous regressors, \textit{i.e.}, endogenous heteroskedasticity. To address the inconsistency of the two-stage least squares estimator in this scenario, and recover the causal parameters of interest, we develop a framework for practical estimation and inference based on the control function approach allowing for discrete and continuous regressors. In particular, we suggest a simple two-step estimation procedure. We establish the limiting properties of the estimator, namely, consistency and asymptotic normality. In addition, we develop practical valid inference methods by proposing an estimator for the asymptotic variance-covariance matrix, and formally establishing its consistency. Monte Carlo simulations provide evidence on the finite-sample performance of the proposed methods and evaluate different implementation strategies. We revisit an empirical application on job training to illustrate the methods.

\end{abstract}


\bigskip


\clearpage
\pagenumbering{arabic} 

\section{Introduction}

This paper studies estimation and inference of linear models with endogeneity and heteroskedasticity. At the estimation stage, endogeneity is typically addressed using instrumental variables (IV) and two-stage least squares (2SLS) procedures. Heteroskedasticity, on the other hand, is usually handled with robust inference techniques. However, when heteroskedasticity is caused by an endogenous regressor, a situation referred to as endogenous heteroskedasticity, the 2SLS estimator may become inconsistent because the necessary exogeneity condition fails. The literature on econometric models that allow for endogenous heteroskedasticity is limited (see, e.g., \cite{Chen08}, \citet{FlorensHeckmanMeghirVytlacil08}, \cite{ChenKhan14}, and \citet{AbrevayaXu23}).

In this paper, we focus on this issue and develop methods for estimation and inference of parameters in linear triangular models with endogeneity and endogenous heteroskedasticity that employ a control function (CF) approach and allow for continuous and discrete regressors. Endogenous heteroskedasticity models can be viewed as a special case of nonseparable models, see, e.g., \cite{Chesher03}, \cite{ChernozhukovImbensNewey07}, \citet{FlorensHeckmanMeghirVytlacil08}, \cite{ImbensNewey09} and \cite{Jun09}. These general methods usually employ nonparametric modeling and estimation together with a CF. Differently from these papers, our interest lies in practical linear regression models in the presence of multiple covariates, such that nonparametric methods become impractical. Thus, we rely on a feasible parametric model using the CF approach.\footnote{There is a large literature on the CF approach. See, e.g., among many others, \citet{NeweyPowellVella99}, \citet{KimPetrin22}, and \cite{han2025setvaluedcontrolfunctions}. Different alternatives have been studied in the literature. See the survey by \cite{BlundellPowell08}, and \cite{ImbensNewey09} for nonparametric and semiparametric models.} We start by building on the general identification results in \citet{FlorensHeckmanMeghirVytlacil08} and tailoring the conditions to accommodate continuous and discrete endogenous variables, as well as exogenous covariates, such that we are able to write the parameter of interest as an explicit function of observables, and thus derive the estimator of interest.
The key restrictions for identification are a CF condition with a flexible parameterization, along with another parameterization of the skedastic function in the structural equation. 
Given a valid CF constructed from the first stage, and the structure of the skedastic function, the parameter of interest in the second stage can be identified using a standard ordinary least squares (OLS) regression. We highlight that our proposed approach includes the 2SLS as a special case when there is no endogenous heteroskedasticity and there is no correction for the first-stage heteroskedasticity, a condition that depends on the CF assumption.
  
Estimation is implemented in a two-step procedure. First, the skedastic function in the first step can be estimated using flexible scale function methods suggested in \citet{RomanoWolf17}. Given the estimate of the skedastic function, the CF can be computed as the normalized errors from the first-stage regression.
In the second step, one performs a simple linear regression with a CF approach. Empirically, this step uses an OLS regression of the dependent variable on the endogenous, exogenous, and CF variables augmented with their interacting terms. We establish the limiting statistical properties of the proposed two-step estimator. Mild sufficient conditions are provided for the estimator to have desired asymptotic properties, namely, consistency and asymptotic normality. The limiting theory presented here incorporates elements of both the CF and generated regressors.

In addition, we develop practical statistical inference procedures by suggesting an estimator for the asymptotic variance-covariance matrix with the generated regressors. We formally establish the consistency of the variance-covariance matrix estimator. The proposed methods provide foundation for general inference procedures. Testing for general linear hypotheses is easily accommodated by Wald type tests, and non-linear hypotheses are readily available by employing the Delta method. We also suggest a test for endogenous heteroskedasticity.

We evaluate the finite sample performance of the proposed methods using Monte Carlo exercises. Numerical simulations confirm that the presence of endogenous heteroskedasticity in the structural equation induces bias in the 2SLS estimator. Moreover, heteroskedasticity in the first-stage equation may amplify the bias. The proposed CF method is able to produce estimates that concentrate around the true value of the target parameter. In addition, results improve when the sample size increases. 

Finally, we present an empirical application that illustrates the framework discussed in this paper. We apply the estimator to study the effect of public-sponsored training program Job Training Partnership Act (JTPA) on future earnings. We estimate the effect of interest using the CF, and for comparison, the 2SLS. Empirical results show evidence that the 2SLS-IV estimator (the typical application in other papers that used the JTPA data) is substantially smaller than the alternative CF methods.

The literature considering the effects of endogenous heteroskedasticity is restricted. 
\citet{Chen08} studies a wage model in which the variance of potential wages depends on the endogenously chosen level of schooling, and addresses the resulting selection with a parametric, Heckman-type control-function approach.
\cite{FlorensHeckmanMeghirVytlacil08} use the CF approach to identify the average treatment effect and the effect of treatment on the treated in models with a continuous endogenous regressor allowing for endogenous heteroskedasticity. They consider a nonparametric model for the endogenous variable and a flexible polynomial for the skedastic function. However, their focus is on the case of no additional exogenous covariates, and estimation and inference are not provided.
\citet{ChenKhan14} discuss identification and estimation of models with endogenous heteroskedasticity in binary treatments using semiparametric IV and pairwise matching. Their main object of interest is different from ours, as they focus on the variance of potential outcomes.
\citet{AbrevayaXu23} show that when a binary treatment is endogenous and there is endogenous heteroskedasticity, that is, the endogenous treatment indicator also affects the scale of the outcome variable, then standard methods for estimation of average treatment effects fail. In particular, they consider the standard IV estimator with binary treatment and show that it is an inconsistent estimator for the average treatment effect. After nonparametric identification is established, a nonparametric IV estimator is provided for the mean and variance treatment effect, as well as the average treatment effect on the treated.\footnote{
Heteroskedasticity can sometimes aid in addressing endogeneity problems rather than exacerbating them, as shown in \citet{Rigobon03}, \citet{KleinVella10}, and \citet{Lewbell12}. However, this is not the case considered in the present study. Our proposed methods, on the other hand, are designed for both continuous or discrete treatment and instrument variables.} 

The methods provided in this paper are complementary to the above approaches. In contrast to the papers listed above, we do not require particular support restrictions for the endogenous variables nor the IV estimation procedure. That is, we do not require continuity as in \citet{FlorensHeckmanMeghirVytlacil08}, nor discreteness as in \citet{AbrevayaXu23} and \cite{ChenKhan14}. Our estimator is deliberately linear and parametric in the covariates, and by using the CF variable it can be simply implemented with standard least-squares regressions.

This paper is organized as follows. Section \ref{sec: model} presents the model and shows the inconsistency of 2SLS under endogenous heteroskedasticity. In Section \ref{sec: CF} we propose a CF approach and discuss identification. Section \ref{sec:examples} illustrates the proposed econometric model with two economic examples.
Estimation and inference procedures are detailed in Section \ref{sec:estimation}. A Monte Carlo study is provided in Section \ref{sec:MC}. Section \ref{sec: Application} illustrates the methods with an empirical application. Finally, Section \ref{sec:Conclusion} concludes. Mathematical proofs are relegated to the Appendix.

\section{A linear model with endogenous heteroskedasticity}\label{sec: model}

In this section, we first describe the model of interest, which contains endogenous heteroskedasticity. Second, we motivate the proposed methods by reviewing the inconsistency of the two-stage least squares in this scenario.

\subsection{Model}

We consider the following model:
\begin{align}
Y&=D\alpha_1 + X'\alpha_2 + g(D, X)\varepsilon \label{eq:structural0} \\ 
D &=Z'\pi_1 + X'\pi_2 + h(X, Z)V,\label{eq:first stage0}
\end{align}
where the main parameter of interest is the scalar \(\alpha_1\). The vector of exogenous variables \(X\) is \((d_x - 1)\)-dimensional (including a constant), and the instrumental variable (IV) \(Z\) is \(d_z\)-dimensional. The variables \(\varepsilon\) and \(V\) are unobservable and their correlation makes \(D\) potentially endogenous.  

The model in equations \eqref{eq:structural0} and \eqref{eq:first stage0} has several important features. Despite the linearity in $D$ in \eqref{eq:structural0}, it allows for heterogeneous effects on $Y$ due to the term $g(D, X)\varepsilon$. Because of this, the parameter of interest has an interpretation as an average effect. Indeed, if we define the potential outcome for unit $i$ when $D=d$ as $Y_{di} = d\alpha_1 + X_i'\alpha_2 + g(d, X_i)\varepsilon_i$, then $E[Y_{di} - Y_{d'i}] = (d - d')\alpha_1$.\footnote{See Appendix \ref{sec: ap interpretation} for more about the interpretation of $\alpha_1$.} 
However, the exogeneity conditions \(E[\varepsilon | X, Z] = 0\) and \(E[V | X, Z] = 0\) are not sufficient to identify \(\alpha_1\) via 2SLS, since \(E[g(D, X)\varepsilon | X, Z]\) may not be zero. Since the conditional variance of the error term in \(\eqref{eq:structural0}\) is \(g(D, X)^2 Var[\varepsilon|D,X]\), we refer to it as ``endogenous heteroskedasticity,'' a terminology borrowed from \cite{AbrevayaXu23}.

The model above is a particular case of non-additive triangular models, where the interest lies in estimating the average linear effect, $\alpha_1$. This specification would allow us to study the role of the skedastic function $g$ in the 2SLS bias for $\alpha_1$ and specific solutions using the control function approach. A comparable model is used in \citet{AbrevayaXu23} to study the role of endogenous heteroskedasticity in treatment effects models. Furthermore, equation \eqref{eq:first stage0} could be rewritten as a general model $D=m(Z,X,V)$ but we use a similar skedastic structure to match equation \eqref{eq:structural0}. The parametric model in the first stage is mainly used due to the curse of dimensionality problem, since most empirical work uses multiple control variables. In a nonparametric case, estimation could rely on methods suggested by \citet{JiSuXiao15} and \cite{LintonXiao19}.

\subsection{Motivation: Inconsistency of the 2SLS estimator}\label{sec: bias2SLS}

The 2SLS is a popular method to address endogeneity in linear models. In this section, we show that the 2SLS estimator estimates the average effect $\alpha_1$ with a bias in the presence of endogenous heteroskedasticity. Although the result here is simple, intuitive, and similar issues have been highlighted in the literature, we include this section to motivate the linear representation above.

The failure of 2SLS to identify $\alpha_1$ in the presence of endogenous heteroskedasticity was highlighted by \cite{AbrevayaXu23} in the case of a binary \(D\). When a binary IV is available, they offer a detailed analysis of the bias and propose a novel identification and estimation strategy. In a similar context with a continuous \(D\), \cite{FlorensHeckmanMeghirVytlacil08} argue that the usual exogeneity IV requirements are not sufficient for identification. 

The following example illustrates the bias in the 2SLS procedure in a model described by \eqref{eq:structural0}--\eqref{eq:first stage0}.
\begin{example}\label{example:locsca general}
Consider the following simple linear case of model \eqref{eq:structural0}--\eqref{eq:first stage0} with one endogenous variable, one instrument, $g(D)=1+\delta D$ and $h(Z)=1+Z\gamma$:
\begin{align*}
Y &=D\alpha_1 + (1+D\delta ) \varepsilon\\  
D &=Z\pi_1 + (1+Z\gamma)V.  
\end{align*}
Here, $E[D\varepsilon]\neq0$ through the possible correlation between $\varepsilon$ and $V$. We assume that  $E[\varepsilon|Z]=E[V|Z]=0$. The 2SLS estimator of $\alpha_1$ is the sample counterpart of $\frac{ Cov[Z,Y]}{ Cov[Z,D]}$. Hence, we have that
\begin{align*}
Cov[Z,Y] &=\alpha_1Cov[Z,D] + Cov[Z,\varepsilon] + \delta Cov[Z, D\varepsilon].
\end{align*}
Now, $E[\varepsilon|Z]=0$ implies $Cov[Z,\varepsilon] =0$. Therefore, we have
\begin{align*}
\frac{ Cov[Z,Y]}{ Cov[Z,D]} &=\alpha_1 + \delta \frac{Cov[Z, D\varepsilon]}{ Cov[Z,D]}.
\end{align*}
The exogeneity assumptions on $Z$ do not imply that $Cov[Z, D\varepsilon]=0$. Therefore, $\alpha_1$ is not identified by the 2SLS moment conditions whenever there is heteroskedasticity in the structural equation, \textit{i.e.,} $\delta\neq 0.$ 

Under the stronger independence assumption: $Z\indep (V,\varepsilon)$ (typical in experimental settings when $Z$ is randomly assigned), the expression simplifies to 
\begin{align*}
\frac{ Cov[Z,Y]}{ Cov[Z,D]} 
&=\alpha_1 + \delta\gamma \frac{E[V\varepsilon]}{\pi_1}
\end{align*}
In this case, the joint heteroskedasticity together with the endogeneity (in the form of $E[V\varepsilon]\neq 0$) is inducing the bias. Thus, in this case, one could have a situation where there is heteroskedasticity in the structural equation with $\delta \neq 0$, but if there is no heteroskedasticity in the first stage, $\gamma=0$, then 2SLS is able to correctly identify the parameter of interest.

\end{example}

Next, we provide a formal result deriving the asymptotic bias in the 2SLS estimator under the endogenous heteroskedasticity. This is not a novel result \textit{per se}, but we specify it to our particular setting.

We maintain the following assumptions.

\begin{assumption}\label{assumption:ind}
$(i)$ The sample ${(Y_{i},D_{i},X_{i}',Z_i')}_{i=1}^n$ is i.i.d. following the model \eqref{eq:structural0}--\eqref{eq:first stage0}. $(ii)$ The following exogeneity conditions hold: $E[\varepsilon|X,Z]=E[V|X,Z]=0$. 
\end{assumption}

\begin{assumption}\label{assumption:rel}
There is at least one component of $\pi_1$ which is not 0, and the matrices $E\begin{bmatrix}
    ZZ'&ZX'\\
    XZ'&XX'
\end{bmatrix}$ and $E\begin{bmatrix}
    XD&XX'\\
    ZD&ZX'
\end{bmatrix}$
have full column rank.
\end{assumption}

Assumptions \ref{assumption:ind} and \ref{assumption:rel} are standard in the literature allowing for endogeneity of $D$, the presence of observable controls $X$, as well as for the valid instruments $Z$ to be (\textit{intrinsically}) exogenous. While the instruments $Z$ are uncorrelated with $\varepsilon$ and $V$, we will see that when we consider the unobservable in \eqref{eq:structural0} to be $g(D, X)\varepsilon$, then it might no longer be the case that $E[Zg(D, X)\varepsilon]= 0$. 
That is, despite $Z$ being uncorrelated to the unobserved innovations, and being relevant, the heteroskedastic nature of the triangular system can invalidate the 2SLS procedure. As an alternative, we propose a control function approach in Section \ref{sec: CF} below. 

The next result compares the probability limit of the 2SLS estimator, denoted by $\overline{\alpha}_{1}$, to the target parameter $\alpha_1$. We assume that the data is generated according to the model in \eqref{eq:structural0}--\eqref{eq:first stage0}, and a researcher who is interested in $\alpha_1$, instruments $D$ with $Z$ in a 2SLS regression of $Y$ on $D$ and $X$ (which includes a constant). In order to deal with the presence of the additional controls $X$, we use a population version of the Frisch-Waugh-Lovell (FWL) theorem using linear projections.\footnote{See Section \ref{sec:app_lin_proj} in the Appendix for a brief review of linear projections.}

\begin{lemma}\label{lemma:bias_2sls}
Consider the model in \eqref{eq:structural0}--\eqref{eq:first stage0}.
Let $\overline{\alpha}_{1}$ be the probability limit of the 2SLS estimator of $\alpha_1$ of a 2SLS regression of $Y$ on $D$ and $X$, with $D$ instrumented by $Z$ and $X$. Under Assumptions \ref{assumption:ind} and \ref{assumption:rel}, we have 
\begin{equation}\label{eq:bias 2sls}
\overline{\alpha}_{1} = \alpha_1 + \Sigma_{h}^{-1} \pi_1'  E[\overline Zg(Z'\pi_1 + X'\pi_2+ h(X,Z)V, X)\varepsilon],
\end{equation} 
where $\Sigma_h:=\pi_1'E[\overline Z\, \overline Z']\pi_1$, and $\overline Z' := Z'-X'E[XX']^{-1}E[XZ']$.
\end{lemma}

We provide some further remarks to understand the intuition and nature of the result in Lemma \ref{lemma:bias_2sls}.
\begin{enumerate}
    \item The asymptotic bias of the 2SLS estimator in equation \eqref{eq:bias 2sls} can be succinctly written as $\Sigma_{h}^{-1} \pi_1'  E[\overline Zg(D, X)\varepsilon]$, but this might obscure the fact that $h(\cdot)$ is also playing a role through $D$, and could, in principle, amplify the bias.
     
     \item If $g(\cdot)\equiv 1$, i.e., there is no heteroskedasticity in the structural equation, then there is no asymptotic bias in the 2SLS estimator because $E[\overline Z {g(D, X)\varepsilon} ]=E[\overline Z {\varepsilon} ]=0$ by Assumption \ref{assumption:ind}. The lack of bias is regardless of $h(\cdot)$. For a general $g(\cdot)$, the form of $h(\cdot)$ may matter.

     \item If $h(\cdot)\equiv 1$, i.e., there is no heteroskedasticity in the first stage, then the asymptotic bias depends only on the form of $g(\cdot)$.

    \item The strength of the instrumental variable, through the magnitude of $\pi_1$, can alleviate the bias. To see this, suppose that $d_z=1$. Then, the asymptotic bias is inversely related to $\pi_1$:
    \begin{align*}
\overline{\alpha}_{1} = \alpha_1 + \frac{E[\overline Z {g(D, X)\varepsilon}]}{\pi_1E[\overline Z^2]} .
\end{align*}
Even in this simple case, the sign of the bias is not totally obvious.

    \item If it is suspected that  $b_0\leq E[\overline Zg(D, X)\varepsilon]\leq b_1$, then partial identification of $\alpha_1$ is possible. Otherwise, this can be used in a sensitivity analysis.

    \item In this paper, we focus on the 2SLS estimator. Nevertheless, we highlight the same result is valid for a GMM estimator. The main intuition is that, under endogenous heteroskedasticity, the moment condition used for estimation would be misspecified through the function $g(\cdot)$.
\end{enumerate}

\section{A control function approach}\label{sec: CF}

This section discusses sufficient conditions, based on the control function (CF) approach, that allow for identification of the parameters of interest in the model given in equations \eqref{eq:structural0}--\eqref{eq:first stage0}, in the presence of both endogeneity and general heteroskedasticity. In the next section, we will suggest a practical estimator and develop inference procedures. Identification in additively separable models with endogenous heteroskedasticity using a CF strategy was initially proposed by \citet{FlorensHeckmanMeghirVytlacil08}. 
Our analysis builds on their results, with the key distinction being a trade-off between making a stochastic polynomial assumption on the CF in our approach versus allowing for a general nonparametric function in their approach. The advantage of our strategy is reflected in the practicality of the estimator. We will show that, by specializing the CF to be a polynomial, the identification result shows that the parameter of interest can be written as a simple explicit function of the data, which in turn can be used for estimation and inference. Moreover, our approach allows the 2SLS-IV estimator to be nested within our model, and thus being a special case.

We maintain the following assumption.
\begin{assumption}\label{assumption:cf}
Control Function Condition: $E[\varepsilon | D,X,Z] =  E[\varepsilon |V]= r(V)$. 
\end{assumption}

Assumption \ref{assumption:cf} is a standard control function condition, see, e.g.,  
\citet{BlundellPowell08}. This is the same condition as Assumption A-3 in \citet[p.1196]{FlorensHeckmanMeghirVytlacil08}. Under this condition, the conditional expectation of the response variable in equation \eqref{eq:structural0} can be written as 
\begin{align}\label{eq:y_cf}    
    E[Y|D, X, Z] &= D\alpha_1 + X'\alpha_2 + g(D, X)E[\varepsilon|D, X, Z]\notag\\ 
    &= D\alpha_1 + X'\alpha_2 + g(D, X)r( V).
\end{align}

It should be noted that, from the previous display, the skedastic term in the structural equation does not depend only on the control function, but also on $g(\cdot)$. The next example, where we assume that both $D$ and $Z$ are continuous, shows that further assumptions are required to identify $\alpha_1$. 

\begin{example}\label{example:cf_ident}
    Consider the model in Example \ref{example:locsca general} without covariates, and suppose that Assumption \ref{assumption:cf} holds. Then \eqref{eq:y_cf} becomes $E[Y|D,Z] =  D\alpha_1  +  g(D) r(\psi(D,Z))$,  where $ V:=\psi(D,Z)$ is identified from the first stage. Define $f(D,Z):=E[Y|D,Z]$. If $D$ and $Z$ are continuous, consider the following derivatives
\begin{align*}    
    \frac{\partial f}{\partial D}
    &=\alpha_1+g_{D}r+g r_{ V}\psi_{D} \\
    \frac{\partial f}{\partial Z}
    &=g r_{ V}\psi_{Z}, 
\end{align*}
where the derivatives are denoted by subscripts, and we omit the arguments in each function. The left hand-sides are identifiable. By multiplying the second equation of the above display by $\psi_{Z}'(\psi_{Z}\psi_{Z}')^{-1}$, and substituting the result into the first equation, we obtain
\begin{align*}    
    \frac{\partial f}{\partial D}
    &=\alpha_1+g_{D}r+\frac{\partial f}{\partial Z}\psi_{Z}'(\psi_{Z}\psi_{Z}')^{-1}\psi_{D}.
    \end{align*} 
Note that, even though $\psi_{Z}$ and $\psi_{D}$ are identified, one cannot identify $\alpha_1$ because of the additional term $g_{D}r$, that is,
\begin{align}\label{eq:ident}    
    \frac{\partial f}{\partial D}-\frac{\partial f}{\partial Z}\psi_{Z}'(\psi_{Z}\psi_{Z}')^{-1}\psi_{D}
    &=
    \alpha_1+g_{D}r.
\end{align}
While the left-hand side of \eqref{eq:ident}  contains only identified objects, the right-hand side contains three unknowns, $\alpha_{1}$, $g_{D}$ and $r$. 
\end{example}

Assumption \ref{assumption:pol2} below imposes a polynomial functional form on the skedastic function.

\begin{assumption}\label{assumption:pol2}
Polynomial Heteroskedasticity: $g(D,X)$ is a polynomial of degree $k_g$ in $D$ and $X$. That is, $g(D,X)=1+\theta_1^{d}D+X'\theta^{x}_{1}+...+\theta^{d}_{k_{g}}D^{k_{g}}+
X'^{k_{g}}\theta^{x}_{k_{g}}$. 
\end{assumption}

This condition is the same as in \citet{FlorensHeckmanMeghirVytlacil08}, except that here it includes additional exogenous covariates.\footnote{\citet[p.1192]{FlorensHeckmanMeghirVytlacil08} write Assumption \ref{assumption:pol2} directly in their model in equations (1) and (2).} In Assumption \ref{assumption:pol2}, $X'^{k_{g}}$ refers to the element-by-element $k_{g}$ power of $X'$. This could be replaced by any other polynomial that may include interactions. 

Finally, we impose the following polynomial structure on the CF. 

\begin{assumption}\label{assumption:pol}
Polynomial Control Function: $r(V)$ is a polynomial of degree $k_v$ in $ V$ that does not contain a constant. That is $r( V)=\sum_{j=1}^{k_v} \mu_j {V}^j$.
\end{assumption}

This assumption specializes to a linear regression model the general nonparametric condition of \citet[][p.1202]{FlorensHeckmanMeghirVytlacil08}. The role of the polynomial structure is to ensure that the skedastic term $g(D,X)$ and the control function $r(V)$ do not share a common component that would be absorbed into, and hence confounded with, the structural parameters $\alpha_1$ and $\alpha_2$. In our specialization, this requirement takes the form of the linear full-rank condition in Assumption \ref{assumption:rank} below, namely that no nonzero $(a,b)$ satisfies $aD+b'W=0$ almost surely; we discuss this condition, and the role of the no-constant normalization imposed above, in Remarks \ref{remark:identification} and \ref{remark:noconstant}. Moreover, in Section \ref{sec:examples} below, we provide economic examples illustrating the proposed methods and discussing these assumptions further.

Identification of the parameter of interest requires identification of $V$, which in turn requires identification of $h$, which can be achieved up to scale. See, for instance, \cite{RomanoWolf17} and \citet[][ch. 8]{Wooldridge12} for parametric specifications, and \citet{JiSuXiao15} and \citet{LintonXiao19} for nonparametric specifications. The next assumption imposes a normalization on the second moment of the control function, and requires the skedastic function to be positive. 

\begin{assumption}\label{assumption:ident h}
  Either $h\equiv 1$ or $E[V^2 | X,Z]=1$ and $h(X,Z)>0$ almost surely.
\end{assumption}

The next lemma considers the identification of $\alpha_{1}$. It shows that the parameter of interest is identified by a regression of $Y$ on transformations of $(D,X,V)$. Define the linear projection of $D$ onto $W$ at $w$ as $L_{[D|W]}(w) := w' E[WW']^{-1}E[WD]$, where $W$ denotes all the regressors in equation \eqref{eq:cond exp y} except $D$, i.e., this contains $X'$ and the interactions of $\{1,D^s,X^{'s}\}_{s=1,...,k_g}$ with $\{{V}^{j}\}_{j=1,...,k_v}$. Collecting the endogenous regressor and these controls, let $R:=(D,W')'$ denote the full vector of second-stage regressors in \eqref{eq:cond exp y}. We impose the following rank condition.

\begin{assumption}\label{assumption:rank}
Second-stage relevance: the matrix $E[RR']$ is nonsingular, where $R:=(D,W')'$ collects the endogenous regressor $D$ and the controls $W$ from \eqref{eq:cond exp y}.
\end{assumption}

We will discuss Assumption \ref{assumption:rank} in Remarks \ref{remark:identification} and \ref{remark:noconstant} below.

\begin{lemma}\label{lemma:identification}
Consider the regression
\begin{align}\label{eq:cond exp y}  
  E[Y|D, X, Z] 
  & = D\alpha_1 + X'\alpha_2 + \left( 1 + \sum_{s=1}^{k_{g}} (\theta_{s}^{d}D^{s}+X'^{s}\theta_{s}^{x}) \right)  \left( \sum_{j=1}^{k_v} \mu_j{V}^j \right) \nonumber \\ & =D\alpha_1 + W'\alpha_w.
\end{align}
 Under Assumptions \ref{assumption:ind}--\ref{assumption:rank}, $\alpha_1$ is the coefficient of $D$ in \eqref{eq:cond exp y} and is given by
\begin{align}\label{eq:alpha_1}
\alpha_1
&= \frac{E[(D- L_{[D|W]}(W))Y]}{E[(D- L_{[D|W]}(W))^2]}.
\end{align}
\end{lemma}

\begin{remark}\label{remark:identification}
Assumption \ref{assumption:rank} is the second-stage counterpart of the relevance condition in Assumption \ref{assumption:rel}. By the Schur-complement identity,\footnote{For a block matrix $M=\big(\begin{smallmatrix} A & B\\ C & E\end{smallmatrix}\big)$ with $E$ invertible, we can alternatively express $M=\big(\begin{smallmatrix} I & BE^{-1}\\ 0 & I\end{smallmatrix}\big)\big(\begin{smallmatrix} A-BE^{-1}C & 0\\ 0 & E\end{smallmatrix}\big)\big(\begin{smallmatrix} I & 0\\ E^{-1}C & I\end{smallmatrix}\big)$. Since the outer matrices have unit determinants, then $\det M=\det(E)\det(A-BE^{-1}C)$. The quantity $A-BE^{-1}C$ is the so-called \emph{Schur complement} of $E$ in $M$. Applied to $M=E[RR']$ with the scalar block $A=E[D^2]$, $B=E[DW']$, $C=E[WD]$, and $E=E[WW']$, this yields \eqref{eq:schur}, and the resulting Schur complement $E[D^2]-E[DW']E[WW']^{-1}E[WD]$ equals the mean-squared residual $E[(D-L_{[D|W]}(W))^2]$ of the linear projection of $D$ on $W$.}
\begin{align}\label{eq:schur}
\det E[RR']&=\det E[WW']\cdot\left(E[D^2]-E[DW']E[WW']^{-1}E[WD]\right)\notag\\&=\det E[WW']\cdot E[(D-L_{[D|W]}(W))^2],
\end{align}
so nonsingularity of $E[RR']$ is equivalent to $E[WW']$ being nonsingular \emph{together with} $E[(D-L_{[D|W]}(W))^2]>0$. Equivalently, there is no nonzero $(a,b)$ with $aD+b'W=0$ almost surely: the linear term $D$ is not collinear with the polynomial and interaction regressors collected in $W$. This guarantees both that the projections $L_{[D|W]}$ and $L_{[Y|W]}$ are well defined and that the denominator in \eqref{eq:alpha_1} is nonzero. Its connection to the no-constant normalization of Assumption \ref{assumption:pol} is discussed in Remark \ref{remark:noconstant}.
\end{remark}

\begin{remark}\label{remark:noconstant}
The no-constant normalization of Assumption \ref{assumption:pol} is a simple \emph{sufficient} condition that rules out one specific violation of Assumption \ref{assumption:rank}, but it is not equivalent to it. Intuitively, the same covariate cannot appear both in the structural linear function and in the implied control function expanded by the skedastic function. Every $D$-term in $W$ has the form $D^sV^j$ with $j\ge1$, so $W$ contains no pure $D$; a constant in $r(V)$ would break this. On the other hand, Assumption \ref{assumption:rank} may hold \emph{with} a constant in $r(V)$ provided $g$ has no linear term---an alternative normalization to that of Assumption \ref{assumption:pol}. Conversely, Assumption \ref{assumption:rank} can fail even when $r(V)$ has no constant, for reasons unrelated to it: for instance a binary $D$ (so that $D^2=D$ and the powers $D^s$ collapse when $k_g\ge2$), a degenerate $V$ (so that $V,\dots,V^{k_v}$ are linearly dependent), or collinear components within $X$. Assumption \ref{assumption:rank} is thus strictly more general.
\end{remark}

\begin{remark}
Note that if $D$ is a dummy variable, as in the empirical application, then there is no need to consider a polynomial model of it and the regression is simplified. The identification result in Lemma \ref{lemma:identification} above shows that the parameter of interest $\alpha_{1}$ in equation \eqref{eq:alpha_1} can be written as a function of observable data. In the next section, we propose a simple two-step estimator for it.
\end{remark}

\begin{remark}\label{remark:discrete}
When $D$ is discrete, \emph{conditional} on $(X,Z)$, the control function $V$ has finite support. For example, in the binary $D$ case, it takes only two values given $(X,Z)$. However, this does not prevent identification, because the rank condition in Assumption \ref{assumption:rank} concerns the \emph{unconditional} distribution of $V$. Since $V$ is governed by the fitted value $Z'\pi_1+X'\pi_2$, it inherits the support of the instrument and the covariates: a continuously varying $Z$ or $X$ delivers continuous unconditional support even for binary $D$, so that the powers $1,V,\dots,V^{k_v}$ remain linearly independent. If $D$, $Z$, and $X$ are all discrete then $V$ necessarily has finite support unconditionally. If the support of $V$ contains $m$ points, the polynomial order must satisfy $k_v\le m-1$. Note that, on $m$ points, $r(V)=E[\varepsilon\mid V]$ is fitted exactly by polynomial of degree $m-1$. This can be interpreted as the parametric counterpart of the large-support conditions of the nonparametric control-function literature \citep{ImbensNewey09,BlundellPowell08}.
\end{remark}

\section{Illustrative examples}\label{sec:examples}

To illustrate the methods we study in this paper, this section provides two simple examples considering different economic models. The examples help to motivate the proposed econometric methods and required assumptions into different economic models with endogenous heteroskedasticity. The first example, inspired by \cite{FlorensHeckmanMeghirVytlacil08}, relates education (the choice variable) and wages.\footnote{This example is also related to the identification of potential wage distributions by education in \citet{Chen08}.}
The second, inspired by \cite{Kitagawa14}, relates labor (the choice variable) and a production function. In both, it is an optimization problem that generates an endogenous choice variable, while the outcome equation exhibits heteroskedasticity. To keep matters simple, we abstract from exogenous covariates in all cases.

\subsection{Heterogeneous returns to education and linear costs}\label{sec:heted}

As mentioned before, this is a version of the model considered by \cite{FlorensHeckmanMeghirVytlacil08}. Suppose that the economic agent receives wages, $Y_{d}$, for given years of education $d$, and wages are given by
\begin{align}
Y_d = \alpha_2 + \alpha_1 d +\frac{\phi_2}{2}d^2 + d\varepsilon ,\label{eq:toy_wage}
\end{align}
where $\varepsilon$ is the unobserved heterogeneity in the
wage level. The endogenous heteroskedasticity term is $d\varepsilon$: two individuals with the same education differ in wages by an amount that is increasing in $d$, so that more education yields not only higher but also more dispersed earnings. Because the agent chooses $d$ optimally after observing $\varepsilon$, education is self-selected on the very heterogeneity that governs its dispersion, so $\varepsilon$ and $d$ are correlated. 

On the other hand, suppose that education costs, $C_d$, are linear in $d$ as follows:
\begin{align}
C_d = c_0 + c_1Z + c_2Zd + d \varepsilon_{1}, \label{eq:toy_cost}
\end{align}
where $\varepsilon_1$ is the unobserved heterogeneity in the cost of schooling.\footnote{The model in \cite{FlorensHeckmanMeghirVytlacil08} has quadratic costs $d^2$ interacted with a function of $Z$. We rule out this case in order to have a linear in $Z$ first stage.} The variable $Z$ is a factor that only affects the cost of schooling, such as the price of education, for example. The agent optimizes the difference, $Y_d-C_d$, by choosing the schooling level $d$. Assuming continuity, the (stochastic) first order condition is
\begin{align}
\frac{\partial (Y_d-C_d)}{\partial d} = \alpha_1 + \phi_2d + \varepsilon  -c_2Z-\varepsilon_1 =0
\label{eq:toy_foc}
\end{align}
and the optimal $d$, provided $\phi_2<0$ (so that the second-order condition for a maximum holds), is
\begin{align}
d = -\frac{\alpha_1}{\phi_2} + \frac{c_2}{\phi_2}Z + \frac{\varepsilon_1-\varepsilon}{\phi_2}. \label{eq:toy_d}
\end{align}
Comparing this optimal choice in \eqref{eq:toy_d} with the first stage \eqref{eq:first stage0}, we have 
\begin{align}\label{eq:toy_firststage_map}
\pi_1=\frac{c_2}{\phi_2},\qquad \pi_2=-\frac{\alpha_1}{\phi_2},\qquad h(X,Z)\equiv 1,\qquad V=\frac{\varepsilon_1-\varepsilon}{\phi_2}.
\end{align}
Thus, the first stage is homoskedastic ($h= 1$), and the control function $V$ is a fixed linear combination of the primitive shocks $(\varepsilon,\varepsilon_1)$. Conditional on $Z$, the optimal choice \eqref{eq:toy_d} is a one-to-one (affine) transformation of $V$. Hence $E[\varepsilon| D,Z]=E[\varepsilon| V,Z]$. Now, if $(\varepsilon,\varepsilon_1)$ is independent of $Z$, then $E[\varepsilon| V,Z]=E[\varepsilon| V]$. This is exactly the control function condition in Assumption \ref{assumption:cf}, with $r(V)=E[\varepsilon\mid V]$.

Substituting the optimal $d$ from \eqref{eq:toy_d} into the wage equation \eqref{eq:toy_wage} yields the realized outcome
\begin{align}\label{eq:toy_structural_map}
Y=\alpha_2+\alpha_1 D+\frac{\phi_2}{2}D^2+D\,\varepsilon,
\end{align}
which is of the form in equation \eqref{eq:structural0} with a single constant covariate ($X=1$, so that $X'\alpha_2=\alpha_2$), skedastic function $g(D)=D$, and augmented by the quadratic term $\tfrac{\phi_2}{2}D^2$. 

Now, for Assumption \ref{assumption:pol} to hold, we need some primitive structure on the error terms. For example, a well-known one is that $(\varepsilon,\varepsilon_1)$ are jointly normal with mean zero.\footnote{Note that the same results can be obtained by directly assuming linearity of the conditional expectation without referring to joint normality. When studying the wage equation under endogenous heteroskedasticity, \citet{Chen08} uses a Heckman-type two-step method that is a special case of the control function approach.} In this case, 
\begin{align}\label{eq:toy_rV}
r(V)=E[\varepsilon| V]=\frac{\operatorname{Cov}(\varepsilon,V)}{\operatorname{Var}(V)} V.
\end{align}
This is a polynomial of degree $k_v=1$ in $V$ with no constant term, which is precisely Assumption \ref{assumption:pol}.

\subsection{Heterogeneous returns to scale in production}\label{sec:toy_production}

We present a second economic model that maps into the econometric model in \eqref{eq:structural0}--\eqref{eq:first stage0}, building on a simple production-function example in \citet[][Section 1]{Kitagawa14}, which is itself based on \citet{MarschakAndrews44}. This example shows that a single productivity shock to the output elasticity of labor is what separates the homoskedastic benchmark, where 2SLS is consistent, from a model with endogenous heteroskedasticity of the form $g(D)=D$, where 2SLS fails.

Consider a firm that produces a homogeneous good, $Q$, using labor $L$, with the Cobb-Douglas technology
\begin{align}\label{eq:kit_tech}
Q(L)=e^{\alpha_2}\,L^{\alpha_1+\varepsilon},
\end{align}
where $\alpha_1$, the output elasticity of labor, is the parameter of interest, $\alpha_2$ is a constant, and $\varepsilon$ is an unobserved mean-zero shock to $\alpha_1$, which satisfies $0<\alpha_1+\varepsilon<1$ almost surely.\footnote{\citet{Kitagawa14} places the unobserved productivity shock on the constant $\alpha_2$ instead: $Q(L)=e^{\alpha_2+\eta}L^{\alpha_1}$. Such a shock enters the log-output equation additively and yields the standard, homoskedastic linear IV model, in which 2SLS is consistent. Placing the shock on the output elasticity $\alpha_1$ instead is what generates the endogenous heteroskedasticity.} Writing $D=\log L$ for the (endogenous) log-input, and $Y=\log Q(L)$ for log-output, the structural equation that \eqref{eq:kit_tech} gives rise to is
\begin{align}\label{eq:kit_structural}
Y=\alpha_1 D+\alpha_2+\varepsilon\,D.
\end{align}

The structural equation \eqref{eq:kit_structural} is exactly of the form \eqref{eq:structural0}, with a single constant covariate ($X=1$, so that $X'\alpha_2=\alpha_2$), skedastic function $g(D)=D$, and structural error $\varepsilon$. The endogenous heteroskedasticity can be interpreted economically as heterogeneity in the elasticity of output with respect to labor. Firms with $\varepsilon>0$ have a higher labor-output elasticity, $\alpha_1+\varepsilon$, and therefore experience a larger percentage increase in output for a given percentage increase in labor input. Note that these differences in firm-specific elasticities reflect unobserved heterogeneity across firms, arising from differences in production technologies, work organization, or other firm-specific factors that affect the responsiveness of output to labor input.

Turning to the firm's behavior, it is assumed that firms face a common output price $p$ and a firm-specific wage $w$. Profits are then $pQ(L)-wL$. Endogeneity arises due to the optimal choice of labor because the firm chooses its input \emph{after} observing its productivity, so $D=\log L$ is correlated with $\varepsilon$.

Solving for the optimal log-labor demand yields
\begin{align}\label{eq:kit_rc_firststage}
D=\frac{\alpha_2+\log(\alpha_1+\varepsilon)+Z}{1-\alpha_1-\varepsilon}.
\end{align}
where $Z:=\log(p/w)$ is the (log) price ratio. This is a linear random coefficients model in the instrument $Z$, and
\begin{align}\label{eq:kit_rc_coef}
D=\gamma_{0}+\gamma_{1}Z,\qquad \gamma_{0}=\frac{\alpha_2+\log(\alpha_1+\varepsilon)}{1-\alpha_1-\varepsilon},\quad \gamma_{1}=\frac{1}{1-\alpha_1-\varepsilon}.
\end{align}

Decomposing each coefficient into its mean and a deviation, $\gamma_{0}=\pi_2+a$ and $\gamma_{1}=\pi_1+b$ with $\pi_2=E[\gamma_{0}]$, $\pi_1=E[\gamma_{1}]$, equation \eqref{eq:kit_rc_coef} becomes
\begin{align}\label{eq:kit_rc_het}
D=\pi_2+\pi_1 Z+U,
\end{align}
where $U:=a+bZ.$ Our parametric first stage \eqref{eq:first stage0} writes the error in the multiplicatively separable form $h(X,Z)V$, with a single scalar $V$. The exact error $U=a+bZ$ in \eqref{eq:kit_rc_het} is not of this form, because $a$ and $b$ are two distinct nonlinear functions of the single shock $\varepsilon$; it instead falls under the general first-stage representation $D=m(Z,X,V)$ noted after \eqref{eq:first stage0}.

In this example the same scalar $\varepsilon$ is both the structural shock and the unobservable driving the first stage: since $D$ is a one-to-one function of $\varepsilon$ given $Z$, the control function may be taken to be $V=\varepsilon$ itself, so
\begin{align}\label{eq:kit_rV}
r(V)=E[\varepsilon\mid V]=E[\varepsilon\mid \varepsilon]=\varepsilon=V.
\end{align}
Assumption \ref{assumption:pol} therefore holds \emph{exactly}, with $r(V)=V$ (that is, $k_v=1$, $\mu_1=1$, and no constant), and \emph{without} the normality assumption required in the wage/cost example: a single-index unobservable makes the control function exact by construction. The price for this cleaner identification is that the first stage is heteroskedastic ($h\not\equiv1$) and only approximately separable. This trade-off is itself informative: the very additivity that makes the standard production-function IV model tractable is what rules out endogenous heteroskedasticity, and relaxing it to obtain a heterogeneous causal effect reintroduces the difficulty that our control-function approach is designed to address.

\section{Estimation and inference}\label{sec:estimation}

This section proposes a practical estimator for the parameter of interest $\alpha_{1}$, and establishes its statistical limiting properties. We consider an estimator that consists of a two-step procedure. The overall procedure is as follows. In the first stage, one estimates the control function (CF), $ V$. In the second-stage, one performs a simple OLS regression of the dependent variable $Y$ on the regressors $(D,X')'$ and the estimated CF, $\widehat{V}$, as well as on their polynomial interactions. In addition, we provide practical inference procedures.

\subsection{First stage estimator: obtaining $\hat V$}\label{section:first_stage}

We employ OLS in the first stage and regress the endogenous variable $D$ on the instruments, $Z$, and exogenous variables $X$, to compute the residuals
\begin{align*}
\check{V}_i &= D_i - Z_i'\widehat{\pi}_{1}  - X_{i}'\widehat{\pi}_{2}.
\end{align*}
Next, we compute the first-stage skedastic function $h(\cdot)$. In practice, to estimate the skedastic function we employ the parametric models in \citet{RomanoWolf17} or \citet[][ch. 8]{Wooldridge12}. For instance,
\begin{equation}\label{scedastic_model_2}
    h(X,Z;\gamma)= \left(|Z'|\gamma_{1}+|X'|\gamma_{2}\right)^{1/2}.
\end{equation}
or 
\begin{equation}\label{scedastic_model_1}
    h(X,Z;\gamma)= \left(\exp\left(|Z'|\gamma_{1}+|X'|\gamma_{2} \right)\right)^{1/2}.
\end{equation}
The coefficients of these models can be estimated using (possibly non-linear) OLS regressions with $\check{V}^{2}_{i}$ or $ln(\check{V}^{2}_{i})$ as the dependent variable and the absolute value of the instruments and exogenous covariates. The resulting estimator is $\widehat{h}_i= h(X_i,Z_i;\widehat{\gamma})$. Using the estimated skedastic function, $\widehat{h}$, we obtain the normalized errors as
\begin{equation*}
    \widehat{V}_{i}= \check{V}_{i} / \widehat{h}_{i}. 
\end{equation*}

The following assumption imposes a standard regularity condition on the skedastic function determining its smoothness and existence of moments of the conditional heteroskedasticity. 

\begin{assumption}\label{assumption:phi_estimator} Assume that $h(X,Z)=h(X,Z;\gamma)$, a parametric function with parameter $\gamma$.
(i) The map $\gamma\mapsto h(x,z;\gamma)$ is twice differentiable with gradient denoted by $ \nabla_{\gamma}h(x,z;\gamma)$; (ii) $E[\nabla_{\gamma}h(X_i,Z_i;\gamma)\nabla_{\gamma}h(X_i,Z_i;\gamma)'h(X_i,Z_i;\gamma)^2]$ has full rank; (iii) the vectors $Z_i$, $X_i$ and the components of $\nabla_{\gamma}h(X_i,Z_i;\gamma)\nabla_{\gamma}h(X_i,Z_i;\gamma)'h(X_i,Z_i;\gamma)^2$ have finite second moments; and (iv) the vectors $Z_ih(X_i,Z_i)V_i$, $X_ih(X_i,Z_i)V_i$ and $ \nabla_{\gamma}h(X_i,Z_i;\gamma) \left(V_i^2-1\right)h(X_i,Z_i;\gamma)^3$ have finite second moments.
\end{assumption}

Denote $\phi:=(\pi_1', \pi_2', \gamma')'$. The next result is the asymptotic normality of $\widehat \phi=(\hat\pi_1', \hat\pi_2', \hat\gamma')'$. The precise form of the influence function $\psi_{\phi,i}$ is detailed in the appendix.

\begin{proposition}\label{prop:phi_if}
Under Assumptions \ref{assumption:ind}, \ref{assumption:rel}, \ref{assumption:ident h}, and \ref{assumption:phi_estimator}, 
\begin{align*}
\sqrt n(\widehat \phi-\phi)=\frac{1}{\sqrt n}\sum_{i=1}^n\psi_{\phi,i}+o_p(1)\overset{d}{\to}\mathcal{N}(0,\Omega_\phi),
\end{align*}
where $\Omega_\phi=E[\psi_{\phi}\psi_{\phi}']$.
\end{proposition}

\subsection{Second stage estimator: augmented control function estimator}\label{section:augmented_control}

In the second step, the goal is to estimate $\alpha_1$ given in equation \eqref{eq:alpha_1}. Recall that from the identification results discussed in the previous section, the population parameter is given by
\begin{align*}
\alpha_1 =\frac{E[(D- L_{[D|W]}(W))Y]}{E[(D- L_{[D|W]}(W))^2]},
\end{align*}
where $W$ is defined as all the regressors on the right hand side of conditional average equation in \eqref{eq:cond exp y}, but $D$.

Denote the preliminary estimator of $ V$ by $\widehat{V}$. The estimator is given by 
\begin{align*}
\widehat{\alpha}_1 =\frac{\sum\limits_{i=1}^n(D_{i}-\widehat{L}_{[D|\widehat{W}]}(W_i) )Y_{i}}{\sum\limits_{i=1}^n(D_{i}-\widehat{L}_{[D|\widehat{W}]}(W_i) ) ^2}.
\end{align*}
where $\widehat{L}[D|\widehat{W}_{i}] = \widehat{W}_{i}'(\widehat{W}'\widehat{W})^{-1}\widehat{W}' D.$ Here $\widehat{W}_{i}$ denotes the vector $W_{i}$ with $ V_i$ replaced by $\widehat{V}_i$. For notational brevity, we write equation \eqref{eq:cond exp y} as
\begin{equation}\label{eq:Y_DW}
  Y  = D\alpha_1 + W(\phi)'\alpha_w + U,
\end{equation}
where $W(\phi)$ contains all the regressors except $D$ as given in Lemma \ref{lemma:identification}, and by construction, $E[U|D,W]=0$. Furthermore, $\alpha=(\alpha_1,\alpha_w')'$.

The next theorem contains the limiting distribution of the proposed two-step CF estimator of $\alpha=(\alpha_1,\alpha_w')'$, denoted $\widehat\alpha=(\widehat\alpha_1,\widehat\alpha_w')'$. This provides details for the asymptotics for the estimator of interest,  \textit{i.e.}, $\alpha_1$, which is an application of OLS with generated regressors. We consider the following assumption:

\begin{assumption}\label{assumption:alpha_1_estimator}
Let $W(\phi)$ be all the regressors in \eqref{eq:Y_DW} (except $D$) as given in Lemma \ref{lemma:identification}, and define $R_i(\phi):=(D_i,W_i(\phi)')'$. Assume that uniformly in $\phi$ (i) $E[R_i(\phi)R_i'(\phi)]$ is non-singular; (ii) a (uniform) law of large numbers holds for $\frac{1}{n}\sum_{i=1}^n R_i(\phi)R_i'(\phi)$, $\frac{1}{n}\sum_{i=1}^n(U_iJ_{\phi,i}- R_i(\phi)\alpha'J_{\phi,i})$, where $J_\phi$ is the Jacobian of $\phi\mapsto R(\phi)$, $J_\phi=\nabla_\phi R(\phi)$; and (iii) $E[\|RR'U^2\|]<\infty$. 
\end{assumption}

Assumption \ref{assumption:alpha_1_estimator} $(i)$ ensures that the CF additional terms to the structural equation are not perfectly collinear to the main regressors, and Assumptions \ref{assumption:alpha_1_estimator} $(ii)$ and $(iii)$ are required for the asymptotic representations using the influence functions.\footnote{Sufficient conditions for a uniform law of large numbers can be found in Lemma 2.4 in \cite{NeweyMcFadden1994}. Essentially, it depends on a dominating function for $R(\phi)$ or, alternatively, $h(x,z;\gamma)$.}

\begin{theorem}   
\label{lemma:limiting_distibution}
Under Assumptions \ref{assumption:ind}-\ref{assumption:alpha_1_estimator},
\begin{align*} 
\sqrt n (\widehat \alpha-\alpha) &=  \frac{1}{\sqrt n}\sum_{i=1}^n\psi_{\alpha,i}  + o_p(1)\overset{d}{\to}\mathcal{N}(0, \Omega_\alpha)
\end{align*}
where $\psi_{\alpha,i}:=E[RR']^{-1}R_i U_i + E[RR']^{-1}E[UJ_\phi- R(\phi)\alpha'J_\phi]\psi_{\phi,i} $, and  $\Omega_\alpha=E[\psi_{\alpha}\psi_{\alpha}'].$ 
\end{theorem}
Note that $\psi_{\phi,i}$ is contributing to $\psi_{\alpha,i}$, the influence function of $\hat\alpha$.
From the above result, we obtain $\sqrt n(\widehat \alpha_1 - \alpha_1)\stackrel{d}{\rightarrow} \mathcal{N}(0,\Omega_{\alpha,(1,1)}),$
where $\Omega_{\alpha,(1,1)}$ is the $(1,1)$ element of $\Omega_{\alpha}$.

\subsection{Inference}\label{sec:inference}

Now we turn our attention to inference in the model under endogenous heteroskedasticity, and suggest a practical estimator for the asymptotic variance-covariance matrix. We also formally establish its consistency. Based on the result of Theorem \ref{lemma:limiting_distibution}, for inference purposes, we deem $\hat \alpha_1$ as approximately normal, centered at $\alpha_1$, and with variance given by $\Omega_{\alpha,(1,1)}/n$.

Estimation of $\Omega_{\alpha,(1,1)}$ can be carried out by resorting to the sample counterpart of $\Omega_{\alpha}$ as follows:
\begin{align*}   
\widehat  \Omega_{\alpha} &= \frac{1}{n}\sum_{i=1}^n\hat\psi_{\alpha,i}\hat\psi_{\alpha,i}'.
\end{align*}
with
\begin{align*}
\hat \psi_{\alpha,i}
&=
\left(\frac{1}{n}\sum_{j=1}^n\widehat R_j \widehat R_j'\right)^{-1}\widehat R_i \widehat U_i 
+ 
\left(\frac{1}{n}\sum_{j=1}^n\widehat R_j \widehat R_j'\right)^{-1}
\left(\frac{1}{n}\sum_{j=1}^n\left(\widehat U_j J_{\hat\phi,j}- \widehat R_j\widehat\alpha'J_{\hat\phi,j}\right)\right)\hat \psi_{\phi,i},
\end{align*}
where
\begin{align*} 
\widehat U_i &= Y_i - D_i\widehat\alpha_1 - W_i(\widehat\phi)'\widehat\alpha_w = Y_i-\widehat R_i'\widehat \alpha,\\
\hat\psi_{\phi,i} &= \hat\Sigma_\phi^{-1}\hat m_i, \\
\widehat{\Sigma}_{\phi}&=\frac{1}{n}\sum_{i=1}^n\begin{bmatrix}
   Z_iZ_i' & Z_iX_i' & 0\\
   X_iZ_i' & X_iX_i' & 0\\
   0        & 0        & 2[\nabla_{\gamma}h(X_i,Z_i;\widehat\gamma)\nabla_{\gamma}h(X_i,Z_i;\widehat\gamma)'h(X_i,Z_i;\widehat\gamma)^2]\\
\end{bmatrix},\\
\widehat m_i&=\begin{bmatrix}
    Z_i\left(D_i - Z_i'\widehat\pi_1 - X_i'\widehat\pi_2\right)\\
    X_i\left(D_i - Z_i'\widehat\pi_1 - X_i'\widehat\pi_2\right)\\
   \nabla_{\gamma}h(X_i,Z_i;\widehat\gamma) \left(\hat V_i^2-1\right)h(X_i,Z_i;\widehat\gamma)^3
\end{bmatrix}.
\end{align*}

Next, we state a result establishing the consistency of the above estimator. 

\begin{lemma}\label{lemma_variance}
    If Assumptions \ref{assumption:phi_estimator} and \ref{assumption:alpha_1_estimator} hold, $E[\|R\|^4]<\infty$, $E[\|J_{\phi}\|^4]<\infty$, $E[U^4]<\infty$, $E[\|Z\|^4]<\infty$, $E[\|X\|^4]<\infty$,$E[|h(X,Z;\gamma)V|^4]<\infty$, $c_i(\phi):=\nabla_{\gamma}h(X_i,Z_i;\gamma) \left( V_i^2-1\right)h(X_i,Z_i;\gamma)^3$ is differentiable in $\phi$ with derivative
$\nabla_\phi c_i(\phi)$, $E\big[\|\nabla_\phi c_i(\phi)\|^2\big] < \infty$, and satisfies a uniform law of large numbers, then $\widehat  \Omega_{\alpha}\overset{p}{\to}\Omega_{\alpha}.$
\end{lemma}

Given the result in Lemma \ref{lemma_variance} and Theorem \ref{lemma:limiting_distibution}, inference is standard. A $100(1-\theta)\%$ confidence interval for $\alpha_1$ can be constructed as $\widehat \alpha_1\pm  z_{\theta/2}\sqrt {\widehat  \Omega_{\alpha,(1,1)}/n}$, where $\widehat  \Omega_{\alpha,(1,1)}$ is the $(1,1)$ element of $\widehat  \Omega_{\alpha}$, and $ z_{\theta/2}$ is the corresponding critical value ($\theta/2$ area to the right) from the standard normal distribution.

Moreover, given the asymptotic normality in Theorem \ref{lemma:limiting_distibution}, general hypotheses on the parameter $\alpha_{1}$, or more generally $\alpha$, can be accommodated by simple Wald-type tests. The
Wald process and associated limiting theory provide a natural foundation for the hypothesis $R\alpha =r$ when $r$ is known, with a Chi-square limiting distribution. Non-linear testing can be easily performed by employing the Delta method.

\subsection{Testing for endogenous heteroskedasticity}

In this section we propose a simple test for endogenous heteroskedasticity, that is, a test of whether the skedastic function $g(D,X)$ actually depends on the endogenous regressor $D$. Writing the identifying regression \eqref{eq:cond exp y} as $E[Y|D,X,Z]=D\alpha_1+W'\alpha_w$ as in Lemma \ref{lemma:identification}, the regressor vector $W$ contains, among its components, the interactions $D^sV^j$ for $s=1,\dots,k_g$ and $j=1,\dots,k_v$; denote by
\begin{equation*}
\beta_{sj}=\theta_s^d\mu_j
\end{equation*}
the coefficient on $D^sV^j$.

Endogenous heteroskedasticity is present exactly when $g(D,X)$ depends on $D$, \textit{i.e.}, when $\theta_1^d,\dots,\theta_{k_g}^d$ are not all zero. Since $\beta_{sj}=\theta_s^d\mu_j$ and the control function is nondegenerate (some $\mu_j\neq0$), this is equivalent to not all the interaction coefficients $\beta_{sj}$ being zero. The null hypothesis of no endogenous heteroskedasticity is therefore
\begin{equation}\label{eq:nullWald}
H_{0}:\ \beta_{sj}=0 \quad\text{for all } s=1,\dots,k_g,\ j=1,\dots,k_v,
\end{equation}
a set of linear restrictions on the regression coefficients that can be tested with a standard Wald statistic.

\begin{example}
In the simplest case $k_{g}=k_{v}=1$ there is a single interaction, and the regression is
\begin{align*}
  E[Y|D, X, Z] &=D\alpha_1 + X'\alpha_2 + \mu V  + \beta\, D V + \eta\, X' V,
\end{align*}
with $\beta=\theta^{d}\mu$ the coefficient on $DV$ and $\eta=\theta^{x}\mu$ that on $X'V$. The test of no endogenous heteroskedasticity is then the single-restriction test $H_{0}:\beta=0$: one simply checks whether the term $DV$ belongs in the regression.
\end{example}

Collect the interaction coefficients into the vector $\beta=(\beta_{sj})$, of dimension $q=k_gk_v$, and let $R$ be the selection matrix that extracts $\beta$ from the full coefficient vector $\alpha=(\alpha_1,\alpha_w')'$, so that $H_0$ reads $R\alpha=0$. The Wald statistic is
\begin{equation}
\label{eq:Wald}
\mathcal{W}_{n} = n\,(R \hat{\alpha})'[R \hat{\Omega} R']^{-1}(R \hat{\alpha}),
\end{equation}
where $\hat{\Omega}$ is the consistent variance estimator of Lemma \ref{lemma_variance}.

\begin{corollary}\label{cor:Chi2}
Under $H_{0}$ and the conditions of Theorem \ref{lemma:limiting_distibution} and Lemma \ref{lemma_variance},
\begin{equation*}
\mathcal{W}_{n}\stackrel{d}{\to} \chi_{q}^{2},
\end{equation*}
where $q$ is the number of restrictions.
\end{corollary}
The proof is omitted since it follows directly from the continuous mapping theorem, the asymptotic normality of the estimator in Theorem \ref{lemma:limiting_distibution}, and the consistency of the asymptotic variance estimator in Lemma \ref{lemma_variance}.

\section{Monte Carlo simulations}\label{sec:MC}

This section provides numerical simulations to assess the finite sample performance of the proposed methods. We use the following version of the model in equations \eqref{eq:structural0} and \eqref{eq:first stage0},
\begin{align}
Y &=D\alpha_1 + X\alpha_2 + g(D,X) \varepsilon, \label{eq:structural mc} \\
D &= Z\pi_1 + X\pi_2 +  h(X,Z)V, \label{eq:first stage mc}\\
\varepsilon &= U + \lambda V \label{eq:error term mc}.
\end{align}

This allows for endogeneity and different types of heteroskedasticity in both the first and second stages. The parameter $\lambda$ allows for $D$ being exogenous (i.e., $\lambda=0$) or endogenous (i.e., $\lambda\neq0$) in the structural equation \eqref{eq:structural mc}. 

We compute simulation results for the cases of $U\sim \mathcal{N}(0,1)$, $V\sim \mathcal{N}(0,1)$, $Z$ following a folded normal distribution, that is, $Z\sim |\mathcal{N}(0,1)|$. Additionally, they are all independent of each other. We set $X\equiv 1$ so that all models have a constant term. Moreover, we use sample sizes of $n\in\{250,500,1000\}$. The number of replications is set to $2000$.

We consider different types of heteroskedastic models given by
\begin{align}
g(D,X)&= D\delta_1 + D^2\delta_2+X\delta_3 , \label{eq:structural mc g}\\
h^2(X,Z)&= Z\gamma_1+X\gamma_2 . 
\label{eq:structural mc h}
\end{align}
Parameters $(\delta_1,\delta_2)$ allow for heteroskedasticity of the endogenous variable in the structural equation (i.e., $\delta_1\neq0$ or $\delta_2\neq0$) and nonlinear effects (i.e., $\delta_2\neq0$), while the parameter $\delta_{3}$ allows for heteroskedasticity of the exogenous variable, and in this case to have a constant term. Since the model includes a constant, this ensures that the variance is not zero under the absence of endogenous heteroskedasticity.
The parameter $\gamma_1$ also allows for $D$ having conditional homoskedasticity (i.e., $\gamma_1=0$) or heteroskedasticity (i.e., $\gamma_1\neq0$) in the first-stage equation \eqref{eq:first stage mc}. 
The parameters are $\alpha_1=\alpha_2=\pi_1=\pi_2=\delta_3=\gamma_2=1$, $\gamma_1\in\{0,1\}$, $\delta_1\in\{0,1\}$ and $\delta_2\in\{0,0.2\}$, thus covering a wide range of scenarios. 

We present results for different estimators: OLS,  standard 2SLS (note that this is equivalent to a standard CF model where $\check V$ is included as an additional regressor in the structural equation), and the proposed CF with one interaction term CF1 $(\widehat{V} ,\widehat{V} D)$, as well as with two interaction terms CF2 $(\widehat{V} ,\widehat{V} D,\widehat{V} D^2)$. 

For the CF models we consider one type of skedastic function  given in \cite{RomanoWolf17}, as given in equation
\eqref{scedastic_model_2},
$h(X,Z;\gamma)=(\gamma_0+|Z|\gamma_{1})^{1/2}$,
such that $\hat{h}=h(X,Z;\hat\gamma)$. 
Let $\check{V}$ be the residuals of the first-stage regression of $D$ on $Z$ and a constant, then $h^2$ is estimated as the regression of  $\check{V}^2$ on $|Z|$ and a constant. Note that this is not the correctly specified skedastic function given in equation \eqref{eq:structural mc h}.

\subsection{Finite sample properties of the endogenous variable parameter estimator}

Tables \ref{table:l1g0} and \ref{table:l1g1} report the simulation results for the case of $\lambda=1$, i.e., endogeneity, for $\gamma_1=0$ and $\gamma_1=1$, respectively. For completeness, Tables \ref{table:l0g0} and \ref{table:l0g1} report the simulation results for the case of $\lambda=0$, i.e., no endogeneity, also for $\gamma_1=0$ and $\gamma_1=1$, respectively. 
In all cases we report empirical bias and variance of the estimators OLS, 2SLS and CF estimators, and for the latter we compute the average of the estimated variance and 95\% empirical coverage using the Gaussian confidence intervals.

The results in Tables \ref{table:l1g0} and \ref{table:l1g1}  clearly indicate that OLS is biased and that 2SLS is also biased in the presence of endogenous heteroskedasticity (i.e., $\delta_1\neq0$ and/or $\delta_2\neq 0$). 
Moreover, note that the parameter $\gamma_1$, i.e., heteroskedasticity in the first stage, aggravates the 2SLS bias. However, the proposed CF estimators provide unbiased results. When $\delta_2=0$, both CF1 and CF2 work, but when $\delta_2=0.2$, only CF2 completely eliminates the bias. Overall, this indicates that the CF approach eliminates the bias, but nonlinear terms may be needed for different functional forms of the heteroskedasticity inducing functions $g$. 

The last two columns of the tables provide the average estimated variance and the 95\% coverage of the CF estimators. In all cases, the estimated variance approximates the simulated variance and it is very similar when $n=1000$. Moreover, the empirical coverage is close to 95\% whenever we have an approximately unbiased estimator.

Tables \ref{table:l0g0} and \ref{table:l0g1} indicate that when there is no endogeneity and no correction is needed from OLS, the CF estimator is unbiased and behaves in a similar fashion to the 2SLS counterpart.

\begin{table}
    \caption{Monte Carlo simulations, $\lambda=1$, $\gamma_1=0$}
    \label{table:l1g0}
    \centering
    \resizebox{\textwidth}{!}{
    \begin{tabular}{ccc|cc|cc|cccc|cccc}
    \hline
    $n$ & $\delta_1$ & $\delta_2$ &
    \multicolumn{2}{c|}{OLS} &
    \multicolumn{2}{c|}{2SLS} &
    \multicolumn{4}{c|}{CF1: $\hat V+\hat V\times D$} &
    \multicolumn{4}{c}{CF2: $\hat V+\hat V\times D +\hat V\times D^2$} \\
    \hline
     & & & Bias & Var. & Bias & Var. & Bias & Var. & Est.Var & Cov.95\% & Bias & Var. & Est.Var & Cov.95\% \\
    \hline
    250  & 0 & 0   & 0.733 & 0.004 & -0.016 & 0.026 & -0.011 & 0.026 & 0.024 & 0.943 & -0.006 & 0.026 & 0.032 & 0.934 \\
    500  & 0 & 0   & 0.734 & 0.002 & -0.006 & 0.011 & -0.004 & 0.011 & 0.020 & 0.957 & -0.006 & 0.012 & 0.012 & 0.946 \\
    1000 & 0 & 0   & 0.735 & 0.001 &  0.000 & 0.005 &  0.001 & 0.005 & 0.006 & 0.958 &  0.003 & 0.006 & 0.006 & 0.942 \\
    250  & 0 & 0.2 & 1.806 & 0.060 &  0.368 & 0.216 &  0.382 & 0.192 & 0.325 & 0.828 & -0.020 & 0.164 & 0.172 & 0.951 \\
    500  & 0 & 0.2 & 1.805 & 0.030 &  0.382 & 0.104 &  0.389 & 0.094 & 0.093 & 0.714 & -0.008 & 0.085 & 0.084 & 0.952 \\
    1000 & 0 & 0.2 & 1.806 & 0.015 &  0.387 & 0.052 &  0.388 & 0.047 & 0.047 & 0.556 & -0.003 & 0.043 & 0.042 & 0.950 \\
    250  & 1 & 0   & 2.056 & 0.080 & -0.037 & 0.337 & -0.018 & 0.308 & 0.300 & 0.945 & -0.032 & 0.299 & 0.336 & 0.950 \\
    500  & 1 & 0   & 2.056 & 0.038 & -0.010 & 0.151 & -0.002 & 0.141 & 0.146 & 0.955 & -0.007 & 0.160 & 0.148 & 0.943 \\
    1000 & 1 & 0   & 2.051 & 0.019 & -0.013 & 0.078 & -0.008 & 0.071 & 0.072 & 0.951 & -0.008 & 0.081 & 0.074 & 0.942 \\
    250  & 1 & 0.2 & 3.131 & 0.216 &  0.372 & 0.777 &  0.403 & 0.693 & 2.236 & 0.894 & -0.024 & 0.645 & 0.702 & 0.946 \\
    500  & 1 & 0.2 & 3.123 & 0.105 &  0.373 & 0.368 &  0.384 & 0.327 & 0.340 & 0.888 & -0.029 & 0.334 & 0.333 & 0.949 \\
    1000 & 1 & 0.2 & 3.122 & 0.054 &  0.384 & 0.182 &  0.392 & 0.162 & 0.170 & 0.832 & -0.025 & 0.166 & 0.165 & 0.946 \\
    \hline
    \end{tabular}
    }
\end{table}

\begin{table}
    \caption{Monte Carlo simulations, $\lambda=1$, $\gamma_1=1$}
    \label{table:l1g1}
    \resizebox{\textwidth}{!}{
    \centering
    \begin{tabular}{ccc|cc|cc|cccc|cccc}
\hline
$n$ & $\delta_1$ & $\delta_2$ &
\multicolumn{2}{c|}{OLS} &
\multicolumn{2}{c|}{2SLS} &
\multicolumn{4}{c|}{CF1: $\hat V+\hat V\times D$} &
\multicolumn{4}{c}{CF2: $\hat V+\hat V\times D +\hat V\times D^2$} \\
\hline
 & & & Bias & Var. & Bias & Var. & Bias & Var. & Est.Var & Cov.95\% & Bias & Var. & Est.Var & Cov.95\% \\
\hline
250  & 0 & 0   & 0.614 & 0.003 & -0.021 & 0.025 & -0.009 & 0.023 & 0.116 & 0.943 & -0.006 & 0.026 & 0.053 & 0.946 \\
500  & 0 & 0   & 0.612 & 0.001 & -0.011 & 0.011 & -0.005 & 0.011 & 0.011 & 0.953 & -0.002 & 0.012 & 0.012 & 0.944 \\
1000 & 0 & 0   & 0.612 & 0.001 & -0.004 & 0.006 & -0.001 & 0.005 & 0.005 & 0.954 & -0.002 & 0.006 & 0.006 & 0.949 \\
250  & 0 & 0.2 & 1.934 & 0.090 &  0.796 & 0.321 &  0.622 & 0.371 & 0.346 & 0.748 & -0.029 & 0.289 & 0.649 & 0.940 \\
500  & 0 & 0.2 & 1.932 & 0.047 &  0.830 & 0.159 &  0.659 & 0.183 & 0.179 & 0.624 & -0.018 & 0.140 & 0.136 & 0.946 \\
1000 & 0 & 0.2 & 1.942 & 0.025 &  0.850 & 0.086 &  0.678 & 0.101 & 0.094 & 0.379 & -0.011 & 0.075 & 0.069 & 0.937 \\
250  & 1 & 0   & 1.835 & 0.066 &  0.305 & 0.337 & -0.038 & 0.397 & 1.295 & 0.946 & -0.044 & 0.406 & 2.308 & 0.939 \\
500  & 1 & 0   & 1.836 & 0.033 &  0.328 & 0.165 & -0.017 & 0.192 & 0.191 & 0.945 & -0.027 & 0.196 & 0.192 & 0.950 \\
1000 & 1 & 0   & 1.834 & 0.017 &  0.340 & 0.082 & -0.017 & 0.097 & 0.094 & 0.946 & -0.023 & 0.093 & 0.095 & 0.954 \\
250  & 1 & 0.2 & 3.147 & 0.271 &  1.135 & 1.059 &  0.620 & 1.185 & 1.071 & 0.864 & -0.097 & 1.062 & 1.057 & 0.940 \\
500  & 1 & 0.2 & 3.167 & 0.149 &  1.182 & 0.487 &  0.663 & 0.553 & 0.564 & 0.838 & -0.060 & 0.526 & 0.491 & 0.942 \\
1000 & 1 & 0.2 & 3.160 & 0.074 &  1.221 & 0.258 &  0.697 & 0.297 & 0.286 & 0.723 &  0.011 & 0.252 & 0.247 & 0.952 \\
\hline
    \end{tabular}
    }
\end{table}

\begin{table}
    \caption{Monte Carlo simulations, $\lambda=0$, $\gamma_1=0$}
    \label{table:l0g0}
    \resizebox{\textwidth}{!}{
    \centering
    \begin{tabular}{ccc|cc|cc|cccc|cccc}
\hline
$n$ & $\delta_1$ & $\delta_2$ &
\multicolumn{2}{c|}{OLS} &
\multicolumn{2}{c|}{2SLS} &
\multicolumn{4}{c|}{CF1: $\hat V+\hat V\times D$} &
\multicolumn{4}{c}{CF2: $\hat V+\hat V\times D +\hat V\times D^2$} \\
\hline
 & & & Bias & Var. & Bias & Var. & Bias & Var. & Est.Var & Cov.95\% & Bias & Var. & Est.Var & Cov.95\% \\
\hline
250  & 0 & 0   & 0.000 & 0.003 &  0.001 & 0.012 &  0.001 & 0.012 & 0.011 & 0.945 &  0.002 & 0.014 & 0.013 & 0.947 \\
500  & 0 & 0   & 0.000 & 0.002 & -0.001 & 0.006 & -0.001 & 0.006 & 0.006 & 0.941 &  0.002 & 0.007 & 0.007 & 0.943 \\
1000 & 0 & 0   & 0.000 & 0.001 & -0.001 & 0.003 & -0.001 & 0.003 & 0.003 & 0.953 &  0.001 & 0.003 & 0.003 & 0.952 \\
250  & 0 & 0.2 & 0.002 & 0.028 & -0.001 & 0.098 & -0.002 & 0.097 & 0.093 & 0.931 & -0.013 & 0.107 & 0.134 & 0.928 \\
500  & 0 & 0.2 & -0.005 & 0.015 & -0.001 & 0.049 & -0.001 & 0.049 & 0.046 & 0.939 &  0.004 & 0.050 & 0.049 & 0.943 \\
1000 & 0 & 0.2 & -0.003 & 0.007 & -0.003 & 0.025 & -0.003 & 0.025 & 0.024 & 0.942 &  0.000 & 0.024 & 0.025 & 0.952 \\
250  & 1 & 0   & -0.002 & 0.036 & -0.010 & 0.152 & -0.009 & 0.151 & 0.148 & 0.940 &  0.026 & 0.179 & 0.315 & 0.936 \\
500  & 1 & 0   &  0.003 & 0.019 &  0.007 & 0.078 &  0.006 & 0.077 & 0.075 & 0.943 &  0.012 & 0.086 & 0.082 & 0.942 \\
1000 & 1 & 0   &  0.000 & 0.010 &  0.006 & 0.040 &  0.006 & 0.040 & 0.037 & 0.936 &  0.005 & 0.043 & 0.041 & 0.951 \\
250  & 1 & 0.2 & -0.006 & 0.103 & -0.009 & 0.379 & -0.010 & 0.376 & 0.364 & 0.939 & -0.007 & 0.389 & 0.369 & 0.937 \\
500  & 1 & 0.2 & -0.004 & 0.051 & -0.010 & 0.188 & -0.011 & 0.187 & 0.176 & 0.944 & -0.014 & 0.199 & 0.189 & 0.939 \\
1000 & 1 & 0.2 &  0.005 & 0.027 & -0.004 & 0.094 & -0.004 & 0.094 & 0.090 & 0.944 & -0.015 & 0.099 & 0.095 & 0.938 \\
\hline
    \end{tabular}
    }
\end{table}

\begin{table}
    \caption{Monte Carlo simulations, $\lambda=0$, $\gamma_1=1$}
    \label{table:l0g1}
    \resizebox{\textwidth}{!}{
    \centering
    \begin{tabular}{ccc|cc|cc|cccc|cccc}
\hline
$n$ & $\delta_1$ & $\delta_2$ &
\multicolumn{2}{c|}{OLS} &
\multicolumn{2}{c|}{2SLS} &
\multicolumn{4}{c|}{CF1: $\hat V+\hat V\times D$} &
\multicolumn{4}{c}{CF2: $\hat V+\hat V\times D +\hat V\times D^2$} \\
\hline
 & & & Bias & Var. & Bias & Var. & Bias & Var. & Est.Var & Cov.95\% & Bias & Var. & Est.Var & Cov.95\% \\
\hline
250  & 0 & 0   & 0.000 & 0.002 &  0.002 & 0.012 &  0.002 & 0.011 & 0.024 & 0.956 &  0.004 & 0.013 & 0.048 & 0.951 \\
500  & 0 & 0   & 0.000 & 0.001 &  0.001 & 0.006 &  0.000 & 0.006 & 0.005 & 0.949 &  0.002 & 0.007 & 0.006 & 0.945 \\
1000 & 0 & 0   & 0.000 & 0.000 &  0.000 & 0.003 &  0.000 & 0.003 & 0.003 & 0.948 &  0.000 & 0.003 & 0.003 & 0.943 \\
250  & 0 & 0.2 & -0.002 & 0.039 & -0.006 & 0.151 & -0.006 & 0.181 & 0.160 & 0.923 &  0.012 & 0.179 & 0.161 & 0.937 \\
500  & 0 & 0.2 & -0.001 & 0.020 & -0.001 & 0.073 & -0.002 & 0.091 & 0.086 & 0.935 &  0.006 & 0.095 & 0.086 & 0.933 \\
1000 & 0 & 0.2 &  0.001 & 0.011 &  0.005 & 0.038 &  0.006 & 0.047 & 0.045 & 0.941 &  0.007 & 0.050 & 0.045 & 0.927 \\
250  & 1 & 0   &  0.003 & 0.034 &  0.008 & 0.177 &  0.011 & 0.215 & 1.039 & 0.937 & -0.002 & 0.228 & 0.217 & 0.943 \\
500  & 1 & 0   & -0.002 & 0.017 & -0.003 & 0.086 & -0.003 & 0.104 & 0.099 & 0.937 & -0.011 & 0.110 & 0.108 & 0.943 \\
1000 & 1 & 0   &  0.001 & 0.009 &  0.003 & 0.043 &  0.005 & 0.051 & 0.050 & 0.940 &  0.006 & 0.054 & 0.054 & 0.948 \\
250  & 1 & 0.2 &  0.000 & 0.120 &  0.004 & 0.494 &  0.000 & 0.603 & 0.567 & 0.936 & -0.010 & 0.637 & 0.713 & 0.926 \\
500  & 1 & 0.2 & -0.006 & 0.061 & -0.013 & 0.244 & -0.016 & 0.301 & 0.291 & 0.944 & -0.004 & 0.339 & 0.304 & 0.931 \\
1000 & 1 & 0.2 &  0.004 & 0.032 &  0.006 & 0.123 &  0.007 & 0.154 & 0.152 & 0.942 & -0.005 & 0.162 & 0.155 & 0.946 \\
\hline
    \end{tabular}
    }
\end{table}

\subsection{Finite sample properties of the test for endogenous heteroskedasticity}

\begin{figure}[htbp]
    \centering
    \includegraphics[width=1.0\textwidth]{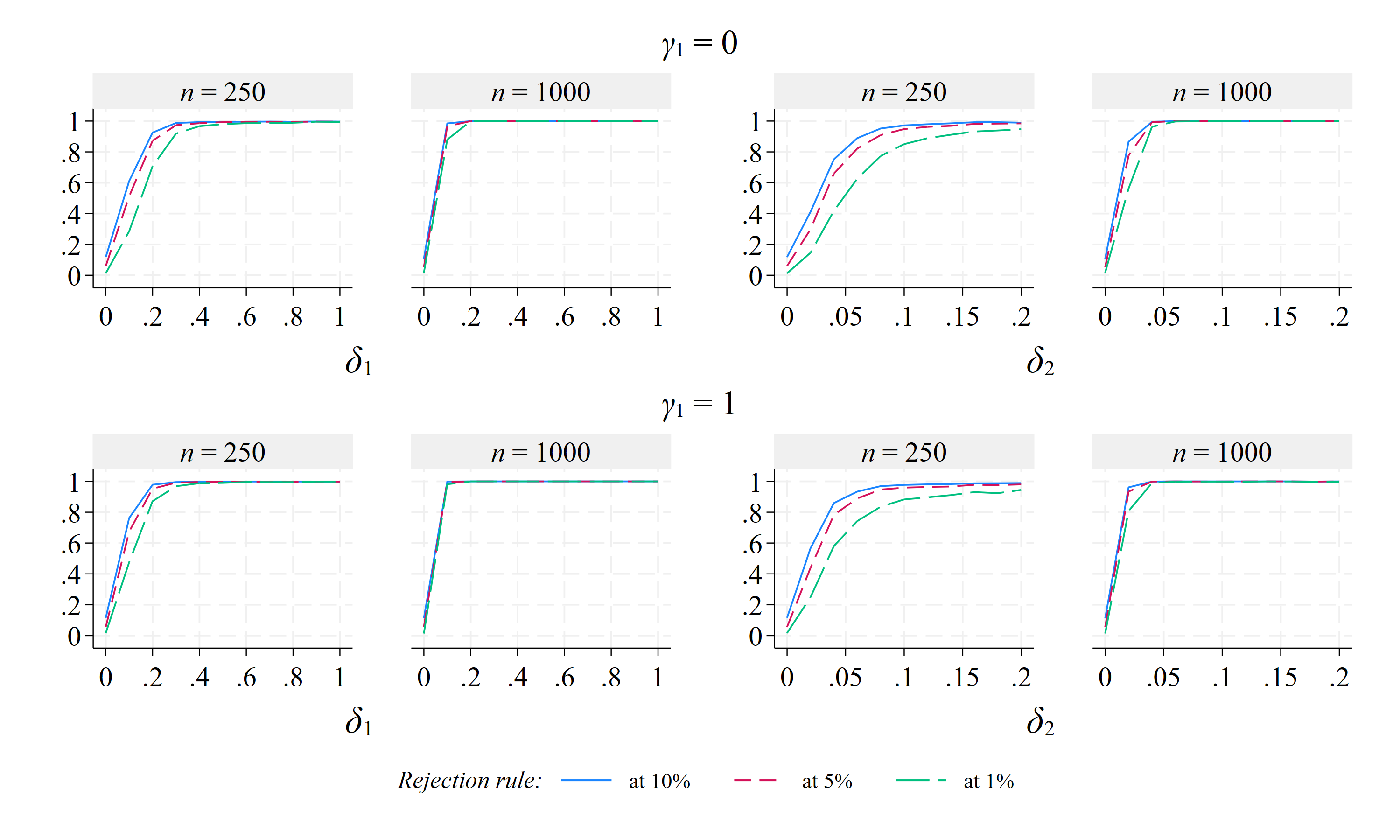}
    \caption{Power of the test using CF1, $\lambda=1$}
    \label{fig:power-cf1}
\end{figure}

\begin{figure}[htbp]
    \centering
    \includegraphics[width=1.0\textwidth]{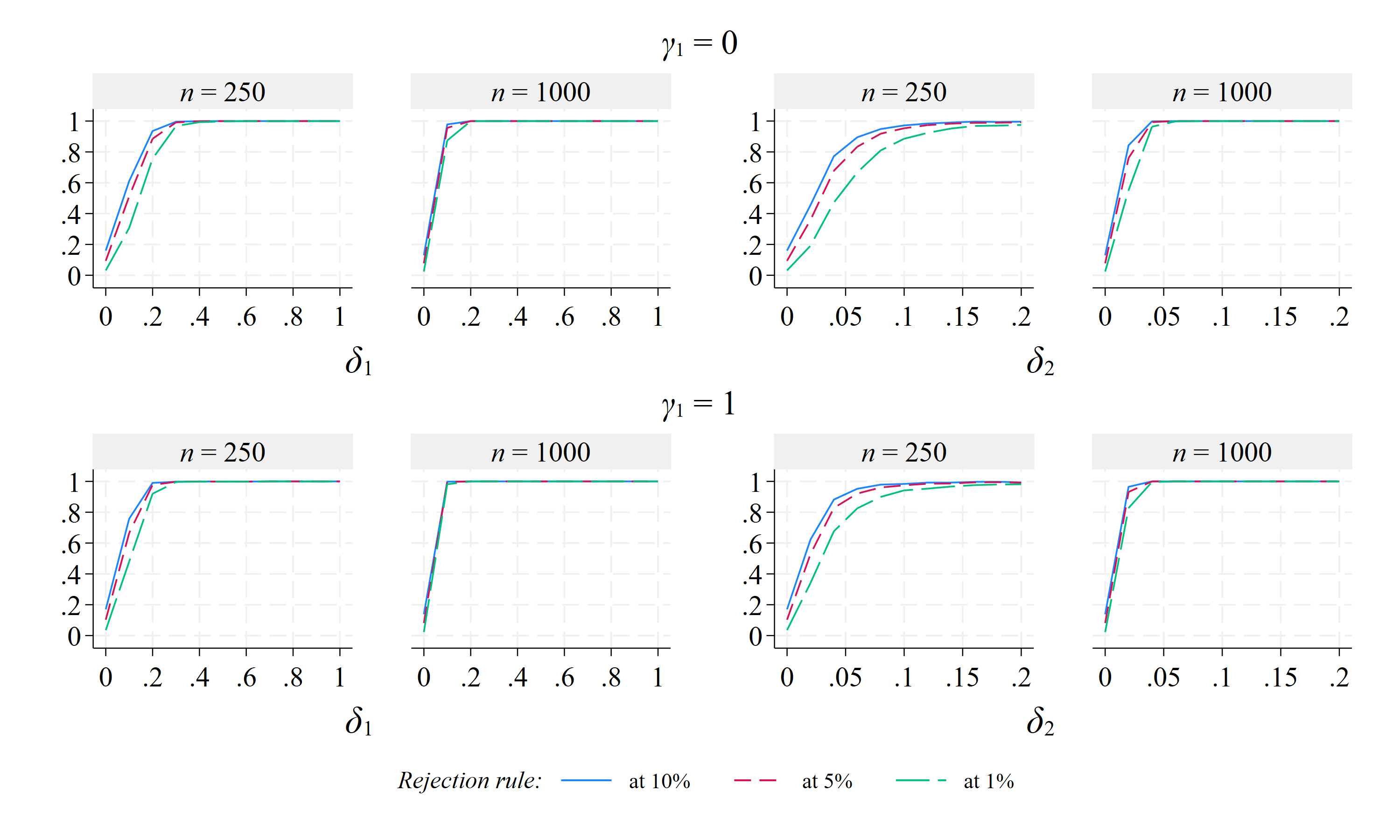}
    \caption{Power of the test using CF2, $\lambda=1$}
    \label{fig:power-cf2}
\end{figure}

Table \ref{table:waldsize} and Figures \ref{fig:power-cf1} and \ref{fig:power-cf2} present the results from implementing the Wald test proposed in Section 5.4 to assess the presence of endogenous heteroskedasticity. Table \ref{table:waldsize} reports the empirical test rejection rates under the null hypothesis, that is, when $\delta_1=\delta_2=0$, for the conventional nominal significance levels of 10\%, 5\%, and 1\%. As it can be seen, when using CF1, empirical rejection rates are close to the nominal levels. In contrast, using a somewhat more complex specification, such as CF2, leads to a slightly oversized test in small samples. Nevertheless, the magnitude of this deviation decreases as the sample size increases, in line with the asymptotic theory of the test.

The parameter $\lambda$ determines the degree of endogeneity in the model, as it controls the correlation between the error term in the outcome equation and the unobserved component of the treatment equation. The results in Table \ref{table:waldsize} show that the behavior of the test in terms of size is similar across different levels of endogeneity, suggesting that the presence of endogeneity does not, by itself, introduce systematic distortions in the rejection rates under the null hypothesis.

Figures \ref{fig:power-cf1} and \ref{fig:power-cf2} show results for the power of the test against different alternative hypotheses, considering separate variations in $\delta_1$ and $\delta_2$, for the endogenous case only ($\lambda=1$). In particular, for each figure, one of these parameters is varied while the other is held fixed at zero. As can be seen, for both CF1 (Figure \ref{fig:power-cf1}) and CF2 (Figure \ref{fig:power-cf2}), the test rejection rate increases sharply as the parameter characterizing the alternative moves away from zero and, consequently, from the null hypothesis of no endogenous heteroskedasticity.

\begin{table}
    \caption{Size of the Wald test for endogenous heteroskedasticity, $H_{0}:\delta_1=\delta_2=0$.}
    \label{table:waldsize}
    \centering
    {\footnotesize
    \begin{tabular}{cccccc|ccc}
\hline
      &       &       & \multicolumn{3}{c|}{CF1: $H_0: \alpha_{VD}=0$} & \multicolumn{3}{c}{CF2: $H_0: \alpha_{VD}=\alpha_{VD^2}=0$} \\
\hline
    $n$     & $\lambda$ & $\gamma_1$ & p<0.10 & p<0.05 & p<0.01 & p<0.10 & p<0.05 & p<0.01 \\
\hline
    250   & 0     & 0     & 0.119 & 0.068 & 0.019 & 0.169 & 0.109 & 0.038 \\
    500   & 0     & 0     & 0.124 & 0.065 & 0.019 & 0.167 & 0.106 & 0.036 \\
    1000  & 0     & 0     & 0.118 & 0.060 & 0.014 & 0.140 & 0.080 & 0.025 \\
    250   & 0     & 1     & 0.117 & 0.064 & 0.021 & 0.175 & 0.110 & 0.038 \\
    500   & 0     & 1     & 0.122 & 0.075 & 0.022 & 0.173 & 0.106 & 0.038 \\
    1000  & 0     & 1     & 0.122 & 0.063 & 0.014 & 0.145 & 0.083 & 0.025 \\
    250   & 1     & 0     & 0.118 & 0.060 & 0.013 & 0.160 & 0.094 & 0.032 \\
    500   & 1     & 0     & 0.106 & 0.059 & 0.013 & 0.159 & 0.093 & 0.026 \\
    1000  & 1     & 0     & 0.108 & 0.053 & 0.017 & 0.130 & 0.078 & 0.024 \\
    250   & 1     & 1     & 0.115 & 0.056 & 0.016 & 0.170 & 0.103 & 0.036 \\
    500   & 1     & 1     & 0.118 & 0.058 & 0.014 & 0.154 & 0.091 & 0.032 \\
    1000  & 1     & 1     & 0.112 & 0.056 & 0.014 & 0.138 & 0.081 & 0.023 \\
\hline
    \end{tabular}%
    }
\end{table}

\subsection{Comparison with the EHIV estimator}

Next, we evaluate results for our estimator relative to the one proposed by \cite{AbrevayaXu23} for the case of binary $D$ and $Z$. We consider the following DGP:
\begin{align*}
Y = D + X + (0.10 + 0.25\,|X| + \delta_1 D)\varepsilon \\
D = 1\left\{\Phi\left[(1+\gamma Z)\eta\right] \geq 0.20\,|X| + 0.50\,Z\right\}, 
\end{align*}
where innovations $(\varepsilon,\eta)$ are bivariate normal random variables with unit variance and correlation $\rho$, $X \sim \mathcal{N}(0,1)$, and $Z$ is a Bernoulli random variable that takes the value one with probability 0.5.  The specification considered by \cite{AbrevayaXu23} corresponds to the case with $\delta_1=\rho=0.5$ and $\gamma=0$.

Note that, since $D$ is binary, heteroskedasticity in the structural equation can only be of a first-order term in $D$. Hence, and in contrast to previous simulations, we only use the CF1 (that augments the regression equation by $\hat V$ and $\hat V D$) to compute our estimator. For the implementation of the nonparametric estimator proposed by \cite{AbrevayaXu23}, we follow their work and use a fourth-order Gaussian kernel $K(u)=0.5(3-u^2)\phi(u)$, $b = 1.06 \times n^{-1/5}$ (Silverman's bandwidth rule) and $\tau = \kappa_0 = \kappa_1 = 0.01$ for the trimming parameters.

We conduct 2000 Monte Carlo simulations using $\rho\in\{0,0.5\}$, $\delta_1\in\{0,0.5\}$, and $\gamma\in\{0,0.5\}$, for $n\in\{250,500,1000\}$. Table \ref{table:ehiv} reports the results for empirical bias and RMSE for the standard 2SLS and CF1, together with the estimator proposed by Abrevaya and Xu, referred to as EHIV (Endogenous Heteroskedasticity Instrumental Variable). Both CF1 and EHIV substantially outperform 2SLS in terms of bias in the presence of endogenous heteroskedasticity, as captured by the parameter configuration $\rho=0.5$ and $\delta_1=0.5$. This difference becomes even more pronounced when heteroskedasticity is also introduced in the first stage ($\gamma=0.5$). Note that EHIV and CF1 exhibit similar behavior in terms of empirical average bias. However, the parametric nature of CF1 provides an efficiency advantage, as reflected in its lower RMSE. Another relevant consideration is that the implementation of EHIV can become computationally more demanding as the number of covariates $X$ increases, due to the curse of dimensionality associated with their nonparametric estimation.

\begin{table}
    \caption{Relative performance against EHIV}
    \label{table:ehiv}
    \centering
    {\footnotesize
    \begin{tabular}{cccc|cc|cc|cc}
\hline
      &       &       &       & \multicolumn{2}{c|}{2SLS} & \multicolumn{2}{c|}{EHIV} & \multicolumn{2}{c}{CF1} \\
\hline
$n$     & $\rho$   & $\delta_1$ & $\gamma$   & Bias  & RMSE  & Bias  & RMSE   & Bias  & RMSE \\
\hline
250   & 0     & 0     & 0     & -0.001 & 0.090 & 0.005 & 0.199 & -0.001 & 0.092 \\
500   & 0     & 0     & 0     & -0.003 & 0.061 & -0.005 & 0.139 & -0.003 & 0.063 \\
1000  & 0     & 0     & 0     & -0.001 & 0.042 & 0.001 & 0.094 & -0.001 & 0.043 \\
\hline
250   & 0.5   & 0     & 0     & -0.005 & 0.085 & -0.016 & 0.210 & -0.007 & 0.087 \\
500   & 0.5   & 0     & 0     & -0.001 & 0.061 & -0.007 & 0.151 & -0.005 & 0.063 \\
1000  & 0.5   & 0     & 0     & -0.001 & 0.043 & -0.005 & 0.111 & -0.005 & 0.044 \\
\hline
250   & 0.5   & 0.5   & 0     & -0.061 & 0.181 & -0.024 & 0.260 & -0.021 & 0.151 \\
500   & 0.5   & 0.5   & 0     & -0.067 & 0.134 & -0.037 & 0.195 & -0.027 & 0.106 \\
1000  & 0.5   & 0.5   & 0     & -0.068 & 0.107 & -0.031 & 0.148 & -0.029 & 0.078 \\
\hline
250   & 0.5   & 0.5   & 0.5   & -0.091 & 0.211 & -0.028 & 0.269 & -0.033 & 0.164 \\
500   & 0.5   & 0.5   & 0.5   & -0.088 & 0.160 & -0.037 & 0.198 & -0.031 & 0.118 \\
1000  & 0.5   & 0.5   & 0.5   & -0.092 & 0.131 & -0.035 & 0.151 & -0.035 & 0.087 \\
\hline
    \end{tabular}%
    }
\end{table}

\section{Empirical application}\label{sec: Application}

We apply the proposed methods to the study of the effect of public-sponsored training programs.\footnote{This application is similar in spirit to Section \ref{sec:heted} for heterogeneous returns to education and endogeneity. Here individuals select into a training program whose earnings consequences are heterogeneous and dispersed, which is a consequence of endogenous heteroskedasticity driven by self-selection.} As argued in \cite{LaLonde95},  public programs of training and employment are designed to improve participants' productive skills, which in turn would increase their earning potential and decrease dependency on social welfare benefits. We use data from the Job Training Partnership Act (JTPA) program, that has been extensively studied in the literature. For example, see \cite{Bloometal97} for a description, and \cite{AbadieAngristImbens02} for applications. The JTPA was a large publicly funded training program that began funding in October 1983 and continued until late 1990s. We focus on the Title II subprogram, which was offered only to individuals with ``barriers to employment'' (long-term use of welfare, being a high-school drop-out, 15 or more recent weeks of unemployment, limited English proficiency, physical or mental disability, reading proficiency below 7th grade level or an arrest record). Individuals in the randomly assigned JTPA treatment group were offered training, while those in the control group were excluded for a period of 18 months. Our interest lies in measuring the effect of actual training on participants' future earnings.

We use the database in \cite{AbadieAngristImbens02} that contains information about adult male and female JTPA
participants and non-participants. Let $Z$ denote the indicator variable for those receiving a JTPA offer. Of
those offered, 60\% completed the training; of those in the control group the completion rate was less than 2\%.  The dependent variable $Y$ is the logarithm of 30-month accumulated earnings (we exclude individuals without earnings), $Z$ is a dummy variable for the JTPA offer, $D$ is the endogenous dummy variable corresponding to the JTPA training, $X$ is a set of exogenous covariates containing individual characteristics. The variables used as detailed in the database are: sex, hsorged, black, hispanic, married, wkless13, afdc, age2225, age2629, age3035, age3644, and age4554. Note that there are many covariates that will not allow us to implement a nonparametric strategy, and this serves as motivation for our parametric approach.
The parameter of interest is the effect of JTPA training on earnings.

\begin{table}[htbp]
  \centering
  \caption{Effect of JTPA training on earnings}
  \label{tab:jtpa}  
    \begin{tabular}{ccccc}
    \toprule
    & Coef. & $earnings$ & $earnings>0$ & $log(earnings)$ \\
    \midrule
OLS & $\alpha_1$    & 2665.7***     & 2052.2***    & 0.2656*** \\
    &                   & (304.7)       & (323.9)   &  (0.0294)\\
    \midrule
2SLS & $\alpha_1$   & 1728.8*** & 1715.7***&  0.1151** \\
     &                     & (498.6) & (518.9) & (0.0485) \\
    & $\pi_1$ & 0.6264***    & 0.6463*** & 0.6463*** \\
    &         & (0.0059)  & (0.0062) & (0.0062) \\
    \midrule
    Control function &       &       &  \\
    \midrule
CF0 & $\alpha_1$       & 2831.9*** & 3071.4** & 0.1652 \\
    &                  & (1065.0) & (1316.3) & (0.1298) \\
    & $\alpha_{\check{V}}$   & 1572.5** & 697.2& 0.2465*** \\
    &             & (643.9) & (686.5) & (0.0675) \\
    & $\alpha_{\check{V}D}$    & -3232.9 & -4215.8 & -0.1557 \\
    &                     & (2818.8) & (3698.7) & (0.3641) \\
    & $\pi_1$            & 0.6264*** & 0.6463*** & 0.6463*** \\
    &                    & (0.0059) & (0.0062) & (0.0062) \\
    \midrule
CF1 & $\alpha_1$        & 2441.9*** & 2106.5*** & 0.1931*** \\
    &                   & (440.0) & (495.8) & (0.0585) \\
    & $\alpha_{\hat{V}}$   & 712.9* & 352.4 & 0.1142*** \\
    &         & (402.1) & (655.1) & (0.0342) \\
    & $\alpha_{\hat{V}D}$    & -999.7* & -711.7 & -0.1172** \\
    &                  & (563.1) & (1126.2) & (0.0448) \\
    & $\pi_1$            & 0.6264*** & 0.6463*** & 0.6463*** \\
    &                    & (0.0059) & (0.0062) & (0.0062) \\
    & $\gamma_1$          & 0.2118*** & 0.2061*** & 0.2061*** \\
    &                    & (0.0024) & (0.0028) & (0.0028) \\
    \bottomrule
    \end{tabular}%

\footnotesize Notes: Standard errors in parentheses. For OLS and 2SLS robust standard errors. CF0: control function with no correction for first-stage skedastic function. CF1: control function with first-stage skedastic function correction. *** p$<$0.01, ** p$<$0.05, * p$<$0.1.\normalsize

\end{table}%

Results for different models appear in Table \ref{tab:jtpa}. Column $earnings$ uses earnings as dependent variable, while $earnings>0$ uses only observations with non-zero earnings, same as the $log(earnings)$ column. Besides OLS and 2SLS, we also consider two control function methods. The first model (CF0) does not consider a skedastic function in the first stage and uses only the first-stage residuals $\check V$. The second model (CF1) applies a first-stage skedastic function using a regression of $\check{V}^2$ on the instrument and a constant ($\check{V}^2=\gamma_0+Z\gamma_{1}+error$ as in the Monte Carlo section) with $\widehat V=\check V/h(X,Z;\hat\gamma)$ and $h(X,Z;\gamma)=(\gamma_0+Z\gamma_{1})^{1/2}$. Both CF models add the estimated $V$ from the first-stage (either $\check V$ for CF0 or $\hat V$ for CF1) and the interaction with $D$ (either $\check V\times D$ for CF0 or $\hat V\times D$ for CF1). All models include other covariates $X$ but coefficients are not reported.

The empirical results in Table \ref{tab:jtpa} show that the OLS point estimates are larger than the 2SLS estimates for all models in levels and logarithms. However, the CF models have $\alpha_1$ estimates closer to OLS but larger than the IV-2SLS estimators. CF0 estimates have very large standard errors for $\alpha_1$ when compared with all other models, and the estimates of $\alpha_{\check{V}D}$ are not statistically significant. Moreover, CF1 has standard errors that are approximately half the CF0 level, and of similar magnitude of 2SLS ones. 

The empirical set-up is similar to that in Example \ref{example:locsca general} for the case of the IV being independent of the error components ($Z$, JTPA offer, is random). For convenience we copy the main result here:
\begin{align*}
\frac{ Cov[Z,Y]}{ Cov[Z,D]} 
&=\alpha_1 + \delta\gamma \frac{E[V\varepsilon]}{\pi_1}=\alpha_1 + 2SLS\ Bias
\end{align*}
Note that the mentioned example suggests that the 2SLS estimate is biased if there is both first-stage ($\gamma$) and second-stage endogenous heteroskedasticity ($\delta$). For the former, the CF1 point estimates of the skedastic first-stage model reveal that $\gamma_1>0$ (i.e., $\gamma$ in the Example) is statistically significant. For the latter, CF1 also shows that $\alpha_{\hat{V}D}<0$ (i.e., $\delta$ in the Example) is also significant, thus indicating endogenous heteroskedasticity. If we assume that $E[V\varepsilon]>0$, i.e., individuals with higher unobservable components in the earnings equation are more likely to select into training, then the 2SLS bias is negative (provided that $\delta<0$, $\gamma>0$, $\pi_1>0$). As such, 2SLS might be downward biased to estimate $\alpha_1$. The CF1 estimates are in line with this argument.

Next, in order to evaluate the estimator's sensitivity to different specifications we take the log earnings model and consider alternative choices. 
We compare different control function models that depend on the specifications of the functions $h$ and $g$. These results appear in Table \ref{tab:jtpa_robust}, where we report the estimated coefficient $\alpha_1$ for the endogenous variable only.
We label different models by columns and rows. Column (A) does not correct for first-stage heteroskedasticity, i.e., $h=1$; column (B) assumes $h(X,Z;\gamma)=(\gamma_0+Z\gamma_1)^{1/2}$ and it is estimated from a regression of $\check{V}^2$ on the instrument and a constant; and column (C) uses $h(X,Z;\gamma)=(\exp(\gamma_0+Z\gamma_1))^{1/2}$, obtained from a regression of $log(\check{V}^2)$ on the instrument and a constant. Second, we compute different skedastic functions $g$ by rows. Here we consider models that augment the regression function by (i) $\hat V$, (ii) $\hat V$ and $\hat V^2$, (iii) $\hat V$ and $\hat V D$, (iv) $\hat V$, $\hat V D$ and $\hat V^2$, and (v) $\hat V$, $\hat V D$, $\hat V^2$ and $\hat V^2D$. 

The results show that the CF  specification in column (A) gives results similar to 2SLS (in fact, by construction (A)-(i) is identical) or not statistically significant. From Table \ref{tab:jtpa} we recall that $\gamma_1$ was statistically significant, and thus an appropriate CF model shall consider first-stage heteroskedasticity. Models in columns (B) and (C) provide comparable results, with (C) giving slightly higher value estimates. The CF1 specification from Table \ref{tab:jtpa}, that corresponds to (B)-(iii) in Table \ref{tab:jtpa_robust}, has estimates comparable to alternative specifications that contain $D$ interacted with powers of $\hat V$ in columns (B) and (C).

\begin{table}[htbp]
\centering
\caption{Robustness analysis of the effect of JTPA training on log earnings}
\label{tab:jtpa_robust}
\begin{tabular}{lccc}
\toprule
OLS
    & 0.266*** & & \\
    & (0.0294) & & \\
\addlinespace
2SLS
    & 0.115** & & \\
    & (0.0485) & & \\
\addlinespace
\midrule
Heter. first-stage & (A) & (B)  & (C) \\
\midrule
\multicolumn{4}{l}{Control function} \\
\addlinespace[2pt]
(i) $\hat V$
    & 0.115** & 0.242*** & 0.263*** \\
    & (0.0485) & (0.0452) & (0.0336) \\
\addlinespace
(ii) $(\hat V,\hat V^2)$
    & 0.147** & 0.186*** & 0.213*** \\
    & (0.0696) & (0.0549) & (0.0462) \\
\addlinespace
(iii) $(\hat V,\hat VD)$
    & 0.165 & 0.193*** & 0.234*** \\
    & (0.1298) & (0.0585) & (0.0379) \\
\addlinespace
(iv) $(\hat V,\hat VD,\hat V^2)$
    & 0.0450 & 0.200*** & 0.227*** \\
    & (0.234) & (0.0555) & (0.0526) \\
\addlinespace
(v) $(\hat V,\hat VD,\hat V^2,\hat V^2D)$
    & -0.604 & 0.180*** & 0.235*** \\
    & (0.463) & (0.0580) & (0.0559) \\
\bottomrule
\end{tabular}

\footnotesize Notes: Standard errors in parentheses. *** p$<$0.01, ** p$<$0.05, * p$<$0.1.\normalsize
\end{table}

Overall, the CF results show empirical evidence that the job training is effective in increasing earnings, but estimates are more modest than those from simple OLS regressions, though larger than 2SLS estimates.

\section{Conclusion}\label{sec:Conclusion}

This paper shows that the estimation of linear models with endogenous heteroskedasticity using the popular 2SLS method may not be estimating the parameter of interest. As an alternative we provide a simple-to-implement augmented control function estimator. The CF approach provides an intuitive modeling strategy to handle endogeneity  and lies in the center of general approaches to this particular set-up. We show that this can be adapted to the least-squares linear estimation methods to correct for the 2SLS bias. In fact, the proposed solution is to augment the standard CF model (which corresponds to 2SLS) with interactions of the structural equation variables with the residuals from the first-stage.

\newpage

\appendix 

\section*{Appendix}

\renewcommand{\thesubsection}{A.\arabic{subsection}}

\setcounter{equation}{0} \renewcommand\theequation{A.\arabic{equation}}

\subsection{Interpretation of $\alpha_1$}\label{sec: ap interpretation}
It is important to attach an interpretation to the parameter $\alpha_1$ in the proposed set-up. To that end, define $y(d,x)$ to be the potential responses of the outcome to different values of the treatment, $d$, and covariates, $x$. Consider the following simple case:
\begin{align*}
y(d,x) = d\alpha_1 + x'\alpha_2.
\end{align*}
First, we discuss the case where $d$ is binary. Then 
\begin{align*}
y(1, x) - y(0, x) = \alpha_1,
\end{align*}
in which case the parameter of interest is the average treatment effect (ATE), because there is no treatment heterogeneity. 

Alternatively, we could define the potential responses to be
\begin{align*}
y(d, x) = d\alpha_1 + x'\alpha_2 + g(d,x)\varepsilon,
\end{align*}
and now we have heterogeneity, because
\begin{align*}
 y(1, x) - y(0, x) = \alpha_1 + (g(1,x) - g(0,x))\varepsilon.
\end{align*}
And then the ATE is
\begin{align*}
E[ y(1, x) - y(0, x)] &= \alpha_1 + (g(1,x) - g(0,x))E[\varepsilon]\\
&= \alpha_1
\end{align*}
if we assume that $E[\varepsilon]=0$. This is similar to the approach taken by \cite{AbrevayaXu23}, where the term $(g(1,x) - g(0,x))\varepsilon$ reflects the variability/deviations from ATE for different individuals/subpopulations. As noted by those authors, the 2SLS bias, i.e., the discrepancy between the ATE and the LATE, depends on ``the degree to which heteroskedasticity depends upon treatment, as well as the average error disturbance for the compliers'' (p.4).

Now, consider the case where $D$ is allowed to take a continuum of values. Now, a more general modeling as in \cite{FlorensHeckmanMeghirVytlacil08} might be useful. Consider the quantity
\begin{align*}
\frac{y(d+\Delta, x) - y(d, x)}{\Delta} = \alpha_1 + \frac{g(d+\Delta,x) - g(d,x)}{\Delta}\varepsilon,
\end{align*}
and the ATE is $\alpha_1$ regardless of $d$, and the size of the jump $\Delta$. If this is the model, then this fits our 2SLS model in the example above.

An alternative is to model in the following way:
\begin{align*}
y(d, x) = \varphi (d) + x'\alpha_2 + g(d,x)\varepsilon
\end{align*}
for some smooth function $\varphi$. Now we have
\begin{align*}
\frac{y(d+\Delta, x) - y(d, x)}{\Delta} = \frac{\varphi(d+\Delta) - \varphi(d)}{\Delta}+ \frac{g(d+\Delta,x) - g(d,x)}{\Delta}\varepsilon,
\end{align*}
and taking the limit we get
\begin{align*}
\lim_{\Delta\to 0}\frac{y(d+\Delta, x) - y(d, x)}{\Delta} = \varphi'(d) + g'(d,x)\varepsilon,
\end{align*}
and the ATE at $d$ is $\varphi'(d)$, while the ``average ATE'' is $E[\varphi'(D)]$. So a linear model 
\begin{align*}
y = E[\varphi'(D)]d + x'\alpha_2 + u
\end{align*}
implies that
\begin{align*}
u = g(d,x)\varepsilon + (\varphi (d) - E[\varphi'(D)]d).
\end{align*}
This is different from our model but it shows that the simple linear representation above can accommodate more general models.

\subsection{Linear projections}\label{sec:app_lin_proj}

Here we briefly review a few properties of linear projections. Let $Y$ be a random variable, and $X$ be a $d_x$-dimensional random vector which we assume always includes a constant. Define the linear projection of $Y$ onto $X$ at $x$ as
\begin{align*}
L_{[Y|X]}(x) = x' E[XX']^{-1}E[XY]. 
\end{align*}
Some properties of $L_{[Y|X]}$ are the following:
\begin{enumerate}
    \item For any constants $a$ and $b$, then $L_{[aY+bW|X]
}(x)=aL_{[Y|X]}(x)+bL_{[W|X]}(x)$.
    \item The residual $\overline Y= Y-L_{[Y|X]}(X)$ is orthogonal to any linear function of $X$ in the following sense: for $\eta\in \mathbb R^{d_x}$ we have
    \begin{align*}
     E[\eta'X\overline Y] &=  E[\eta'X(Y-  L_{[Y|X]}(X))]\\
    &=E[\eta'XY - \eta'XX' E[XX']^{-1}E[XY] ]\\
    &=\eta'E[XY] - \eta' E[XX']E[XX']^{-1}E[XY]\\
    &=\eta'E[XY] - \eta'E[XY]\\
    &=0.
    \end{align*}
\item If $Y$ is $p_y$-dimensional with $p_y>1$, then we define:
\begin{align*}
L_{[Y|X]}(x) =  \underbrace{x'}_{1\times d_x} \underbrace{E[XX']^{-1}}_{d_x\times d_x}\underbrace{E[XY']}_{d_x\times p_y}.
\end{align*}
\end{enumerate}

\subsection{Proof of Lemma \ref{lemma:bias_2sls}}

First, with some abuse of notation, define the matrices (in the usual way) $M_{X}= I-X[X'X]^{-1}X'$ and $P_{M_{X}Z}=M_{X}Z[Z'M_{X}Z]^{-1} Z'M_{X}$. Let $\widehat {\alpha}_{1,2SLS}$ be the 2SLS estimator of a regression of $Y$ on $D$ and $X$, instrumented by $Z$:
 \begin{align*}
     \widehat {\alpha}_{1,2SLS}=(D' P_{M_{X} Z} D)^{-1}  D'P_{M_{X} Z} Y. 
 \end{align*}

The 2SLS is a minimum distance estimator --- see Section 3.8 in \cite{hayashi2011econometrics} --- based on the following moment condition: $E[\overline Z\,\overline Y]=E[\overline Z\,\overline D]\alpha_1$. The population version is the probability limit of $\widehat {\alpha}_{1,2SLS}$, and is given by 
\begin{align}\label{eq:plim_2sls}
\overline \alpha_1 = \left(E[\overline Z\,\overline D]' E[\overline Z\, \overline Z']^{-1} E[\overline Z\,\overline D]\right)^{-1} E[\overline Z\,\overline D]'E[\overline Z\,  \overline Z']^{-1} E[\overline Z\,\overline Y].
\end{align}
Recall that the model in \eqref{eq:structural0}--\eqref{eq:first stage0} is
\begin{align*}
Y&=D\alpha_1 + X'\alpha_2 + g(D, X)\varepsilon, \\ 
D &=Z'\pi_1 + X'\pi_2 + h(X, Z)V.
\end{align*}
Apply the linear projection operator to the structural equation:
\begin{align*}
L_{[Y|X]}(X)&=L_{[D|X]}(X)\alpha_1 + X'\alpha_2 + L_{[g(D, X)\varepsilon|X]}(X).
\end{align*}
The residualized $Y$ is 
\begin{align}\label{eq:y_res}
\overline{Y}&=\overline{D}\alpha_1 + \overline{g\varepsilon},
\end{align}
where $g:=g(D,X)$ for notational simplicity. Plugging \eqref{eq:y_res} into \eqref{eq:plim_2sls}, we obtain
\begin{align}\label{eq:plim_2sls_2}
\overline \alpha_1 = \alpha_1 + \underbrace{\left(E[\overline Z\,\overline D]' E[\overline Z\, \overline Z']^{-1} E[\overline Z\,\overline D]\right)^{-1} E[\overline Z\,\overline D]'E[\overline Z\,  \overline Z']^{-1} E[\overline Z\,\overline{g\varepsilon}]}_{\text{bias}}.
\end{align}
Apply a linear projection to the first-stage:
\begin{align*}
L_{[D|X]}(X)&=L_{[Z'\pi_1|X]}(X) + X'\pi_2 + L_{[h(X, Z)V|X]}(X).
\end{align*}
Therefore, the residualized version is
\begin{align}\label{eq:res_first_stage}
\overline D &=\overline Z'\pi_1  + \overline{hV}.
\end{align}
where $h:=h(X,Z)$ for notational simplicity. Now we plug-in \eqref{eq:res_first_stage} into the bias part of \eqref{eq:plim_2sls_2}. First note that
\begin{align*}
E[\overline Z\,\overline D] &= E[\overline Z\,\overline Z']\pi_1  + E[\overline Z\,\overline{hV}]\\
&=E[\overline Z\,\overline Z']\pi_1,
\end{align*}
because, by Assumption \ref{assumption:ind}, $E[V|X,Z]=0$, which implies $E[\overline Z\,\overline{hV}]=0.$ So that
\begin{align*}
 E[\overline Z\,\overline D]'E[\overline Z\,  \overline Z']^{-1} E[\overline Z\,\overline{g\varepsilon}] &= \pi_1 'E[\overline Z\,\overline Z']  E[\overline Z\,  \overline Z']^{-1} E[\overline Z\,\overline{g\varepsilon}]\\
 &=\pi_1'E[\overline Z\,\overline{g\varepsilon}].
\end{align*}
Now for the other term in the bias
\begin{align*}
E[\overline Z\,\overline D]' E[\overline Z\, \overline Z']^{-1} E[\overline Z\,\overline D] 
&= \pi_1 'E[\overline Z\,\overline Z']E[\overline Z\, \overline Z']^{-1} E[\overline Z\,\overline Z']\pi_1 \\
&=\pi_1 'E[\overline Z\,\overline Z']\pi_1\\
&=:\Sigma_h.
\end{align*}
Putting all together we get that
\begin{align*}
\overline \alpha_1 = \alpha_1 + \Sigma_{h}^{-1}\pi_1'E[\overline Z\,\overline{g\varepsilon}].
\end{align*}
Now, since $g:=g(D, X)$, and $D =Z'\pi_1 + X'\pi_2+ h(X,Z)V,$ the final expression for the asymptotic bias is
\begin{align*}
\text{bias} 
& = \Sigma_{h}^{-1} \pi_1'  E[\overline Zg(Z'\pi_1 + X'\pi_2+ h(X,Z)V, X)\varepsilon].
\end{align*}
\qed

\subsection{Proof of Lemma \ref{lemma:identification}}
Since $h\equiv 1$, $V$ is identified from the first-stage in \eqref{eq:first stage0} as $V = D - Z'\pi_1 - X'\pi_2$. Under Assumptions \ref{assumption:cf}, \ref{assumption:pol2}, and \ref{assumption:pol}, we have the following conditional expectation:
\begin{align}\label{eq:x2_tilde}    
  E[Y|D, X, Z] 
  & = D\alpha_1 + X'\alpha_2 +  \left( 1 + \sum_{s=1}^{k_{g}} (\theta_{s}^{d}D^{s}+X'^{s}\theta_{s}^{x}) \right)  \left( \sum_{j=1}^{k_v} \mu_jV^j \right) . 
\end{align}
Define $W$ to be all the regressors, but $D$, on the right hand side of \eqref{eq:x2_tilde}. Then, we can write
\begin{align}\label{eq:y_fwl}    
  Y & = D\alpha_1 + W'\alpha_w + U,
\end{align}
where $U:= Y- E[Y|D, W] $. Since $W$ is $\sigma(D,X,Z)$-measurable, the tower property gives $E[Y|D,W]=E[E[Y|D,X,Z]\,|\,D,W]=D\alpha_1+W'\alpha_w$, so that $U=Y-E[Y|D,X,Z]$ and hence $E[U|D,X,Z]=0$; in particular $E[UD]=0$ and $E[UW]=0$. The nonsingularity of $E[RR']$, with $R=(D,W')'$, ensures via the Schur-complement identity \eqref{eq:schur} that $E[WW']$ is nonsingular---so the linear projections below are well defined---and that $E[(D-L_{[D|W]}(W))^2]>0$. Now we use linear projections on $W$. By the projection of $Y$ on $W$ we have that $L_{Y|W}(W)  =  L_{[D|W]}(W)\alpha_1 + W'\alpha_w$, since $L_{U|W}(W)=0$ by $E[UW]=0$, so that $ Y - L_{Y|W}(W)  = (D- L_{[D|W]}(W))\alpha_1 + U.$ Multiplying by $D-L_{[D|W]}(W)$ and taking expectations, the term $E[(D-L_{[D|W]}(W))U]=0$ because $D$ and $L_{[D|W]}(W)$ are $\sigma(D,X,Z)$-measurable and $E[U|D,X,Z]=0$; therefore
\begin{align*}
\alpha_1 &= \frac{E[(D- L_{[D|W]}(W))(Y-L_{Y|W}(W))]}{E[(D- L_{[D|W]}(W))^2]}\\
&=\frac{E[(D- L_{[D|W]}(W))Y]}{E[(D- L_{[D|W]}(W))^2]},
\end{align*}
where the last equality uses that the projection residual $D-L_{[D|W]}(W)$ is orthogonal to $W$, and hence to the linear function $L_{Y|W}(W)$ of $W$.
If $h\nequiv 1$, then the first-stage in \eqref{eq:first stage0} is given by $D =Z'\pi_1 + X'\pi_2+ h( X,Z)V$. Consider the residual $h( X,Z)V$, and apply conditional expectations as follows:
\begin{align}\label{ols_fs_1}
E[h(X,Z) V|X,Z]&=h(X,Z)E[V|X,Z]=0,
\end{align}
where $E[V|X,Z]=0$ follows from Assumption \ref{assumption:ind}, and
\begin{align}\label{ols_fs_2}
E[h(X,Z)^2 V^2|X,Z]&=h(X,Z)^2E[V^2|X,Z]=h(X,Z)^2,
\end{align}
which follows from Assumption \ref{assumption:ident h}.
Thus, it is in principle possible to identify $h(X,Z)$ in a two-stage procedure. First, identify $h(X,Z) V$ as the residuals from the regression of $D$ on $Z$ and $X$, which is possible by \eqref{ols_fs_1}. Second, following \eqref{ols_fs_2}, run a regression of $h(X,Z)^2 V^2$ on $X$ and $Z$ to obtain $h(X,Z)^2$. Finally, construct $V$ as
\begin{equation}\label{eq: FS aux}
V = \frac{D - Z'\pi_1 - X'\pi_2}{h(X,Z)}, 
\end{equation}
which is possible since $h(X,Z)\neq 0$ almost surely. Once $V$ is identified, we can proceed as in the case where $h\equiv1$. That is, $\alpha_1$ is identified by \eqref{eq:alpha_1}.

\qed

\subsection{Proof of Proposition \ref{prop:phi_if}}
Here we are going to derive the influence function of $\sqrt n(\widehat \phi-\phi)$. The procedure is detailed in Section \ref{section:first_stage}, and we reproduce it here for completeness. Following Assumption \ref{assumption:phi_estimator}, the first-stage equation is $D =Z'\pi_1 + X'\pi_2+ h(X,Z;\gamma)V.$ By Assumption \ref{assumption:ind}, $E[V|X,Z]=0$, so that $E[D|X,Z] =Z'\pi_1 + X'\pi_2,$ and the OLS regression of $D$ on $Z$ and $X$ yields consistent estimators of $\pi_1$ and $\pi_2$, denoted by $\widehat\pi_1$ and $\widehat\pi_2$ respectively. That is, $\widehat\pi_1$ and $\widehat\pi_2$ satisfy the empirical moment condition:
\begin{align}\label{eq:mom_cond_pi} 
\frac{1}{n}\sum_{i=1}^n\begin{bmatrix}
    Z_i\\X_i
\end{bmatrix} \left(D_i - Z_i'\widehat\pi_1 - X_i'\widehat\pi_2\right)=0.
\end{align}
The residuals of the regression are denoted by $\check V$, and are constructed as $\check {V}_i = D_i - Z_i'\widehat{\pi}_{1}  - X_i'\widehat{\pi}_{2}.$ To estimate $\gamma$, the parameter governing the skedastic function, a model needs to be chosen. Examples are $h(X,Z;\gamma)^2= \exp\left(\gamma_{0}+|Z'|\gamma_{1}+|X'|\gamma_{2} \right)$ and $ h(X,Z;\gamma)^2= \gamma_{0}+|Z'|\gamma_{1}+|X'|\gamma_{2}$. By assumption, $E[V^2 | X,Z]=1$ and $h(X,Z;\gamma)\neq 0$ almost surely, so that 
\begin{align*} 
E[h(X,Z;\gamma)^2V^2|X, Z] = E[(D - Z'\pi_1 - X'\pi_2)^2|X, Z] = h(X,Z;\gamma)^2. 
\end{align*}
This means that a (possibly nonlinear) least squares regression of $(D - Z'\pi_1 - X'\pi_2)^2$ on a transformation of $Z$ and $X$ (given by either \eqref{scedastic_model_2} or \eqref{scedastic_model_1}) consistently estimates $\gamma$.\footnote{A classic reference for nonlinear least squares regression is \cite{jennrich1969}. A modern treatment is given in Example 5.27 of \cite{vanderVaart98}.} This is, of course, infeasible, so we use $\check V^2$. To be more precise, $\widehat\gamma$ is obtained as
\begin{align}\label{eq:gamma_hat_1} 
\widehat\gamma = \arg\min_{a}\sum_{i=1}^n \left(\check V_i^2-h(X_i,Z_i;a)^2\right)^2.
\end{align}
By Assumption \ref{assumption:phi_estimator}, $\gamma\mapsto h(x,z;\gamma)$ is differentiable, then by the first order condition of \eqref{eq:gamma_hat_1}, $\widehat \gamma$ satisfies
\begin{align}\label{eq:mom_cond_gamma}  
\frac{1}{n}\sum_{i=1}^n \nabla_{\gamma}h(X_i,Z_i;\widehat\gamma) \left(\check V_i^2-h(X_i,Z_i;\widehat\gamma)^2\right)h(X_i,Z_i;\widehat\gamma) = 0,
\end{align}
where $\nabla_{\gamma}h(x,z;\widehat\gamma)$ is the gradient of the map $\gamma\mapsto h(x,z;\gamma)$ evaluated at $\gamma = \widehat\gamma$. Combining the empirical moment conditions given in \eqref{eq:mom_cond_pi} and \eqref{eq:mom_cond_gamma} we can write the two-step estimation problem in a single step with $(\widehat\pi_1', \widehat\pi_2', \widehat\gamma')'$ simultaneously satisfying
\begin{align}\label{eq:mom_cond_pi_gamma} 
\frac{1}{n}\sum_{i=1}^n\begin{bmatrix}
    Z_i\left(D_i - Z_i'\widehat\pi_1 - X_i'\widehat\pi_2\right)\\
    X_i\left(D_i - Z_i'\widehat\pi_1 - X_i'\widehat\pi_2\right)\\
    \nabla_{\gamma}h(X_i,Z_i;\widehat\gamma)\left((D_i - Z_i'\widehat{\pi}_{1}  - X_i'\widehat{\pi}_{2})^2-h(X_i,Z_i;\widehat\gamma)^2\right)h(X_i,Z_i;\widehat\gamma)
\end{bmatrix}=0.
\end{align}
Define \begin{align*} 
m_i(\phi):=\begin{bmatrix}
    Z_i\left(D_i - Z_i'\pi_1 - X_i'\pi_2\right)\\
    X_i\left(D_i - Z_i'\pi_1 - X_i'\pi_2\right)\\
   \nabla_{\gamma}h(X_i,Z_i;\gamma) \left(V_i^2-1\right)h(X_i,Z_i;\gamma)^3
\end{bmatrix},
\end{align*}
so that \ref{eq:mom_cond_pi_gamma} can be written as
\begin{align}\label{eq:mom_cond_pi_gamma_2} 
\frac{1}{n}\sum_{i=1}^n m_i(\widehat \phi) = 0.
\end{align}

Next, we do a mean value expansion of the sample moment around the true value $\phi$.  By a first-order Taylor expansion, which is possible by Assumption \ref{assumption:phi_estimator}, 
\begin{align*}
0 
&= \frac{1}{n}\sum_{i=1}^n m_i(\widehat\phi) \\
&= \frac{1}{n}\sum_{i=1}^n m_i(\phi) 
   + \left( \frac{1}{n}\sum_{i=1}^n 
       \frac{\partial m_i(\phi)}{\partial \phi'} 
     \right)
     (\widehat\phi - \phi) + o_p(\|\widehat\phi - \phi\|)\\
&=    \frac{1}{n}\sum_{i=1}^n m_i(\phi) 
   + E\left[ 
       \frac{\partial m_i(\phi)}{\partial \phi'} 
     \right]
     (\widehat\phi - \phi) \\
&+\left( \frac{1}{n}\sum_{i=1}^n 
       \frac{\partial m_i(\phi)}{\partial \phi'} -E\left[ 
       \frac{\partial m_i(\phi)}{\partial \phi'} 
     \right]
     \right)
     (\widehat\phi - \phi)+ o_p(\|\widehat\phi - \phi\|) .
\end{align*}
First, we are going to show that Assumptions \ref{assumption:rel} and \ref{assumption:phi_estimator} imply that 
\begin{align*}
 \frac{1}{n}\sum_{i=1}^n 
       \frac{\partial m_i(\phi)}{\partial \phi'} -E\left[ 
       \frac{\partial m_i(\phi)}{\partial \phi'} 
     \right]=o_p(1),
\end{align*}
and $E\left[ 
       \frac{\partial m_i(\phi)}{\partial \phi'} 
     \right]$ is invertible.
To that end, we compute the derivative. 
\begin{align*}
\frac{\partial m_i(\phi)}{\partial \phi'} = \begin{bmatrix}
 -Z_iZ_i' & -Z_iX_i' & 0\\
   -X_iZ_i' & -X_iX_i' & 0\\
   -2\nabla_{\gamma}h_iV_iZ_i'h_i^2  &-2\nabla_{\gamma}h_iV_iX_i'h_i^2 & \Sigma_\gamma
\end{bmatrix},
\end{align*}
where $h_i:=h(X_i,Z_i;\gamma)$. Note that $$E[-2\nabla_{\gamma}h(X_i,Z_i;\gamma)V_iZ_i'h(X_i,Z_i;\gamma)^2]=E[-2\nabla_{\gamma}h(X_i,Z_i;\gamma)V_iX_i'h(X_i,Z_i;\gamma)^2]=0$$ because by Assumption \ref{assumption:ind}, $E[V|X,Z]=0$. Also note that $E[\nabla_{\gamma}h(X_i,Z_i;\gamma) \left(V_i^2-1\right)h(X_i,Z_i;\gamma)^3]=0$ because we assumed $E[V^2 | X,Z]=1$. The expression for $\Sigma_\gamma$ is
\begin{align*}
\Sigma_\gamma:&=\frac{\partial }{\partial \gamma}\nabla_{\gamma}h(X_i,Z_i;\gamma)\left((D_i - Z_i'\pi_{1}  - X_i'\pi_{2})^2-h(X_i,Z_i;\gamma)^2\right)h(X_i,Z_i;\gamma)\\
&=\nabla^2_{\gamma\gamma}h(X_i,Z_i;\gamma)\left((D_i - Z_i'\pi_{1}  - X_i'\pi_{2})^2-h(X_i,Z_i;\gamma)^2\right)h(X_i,Z_i;\gamma)\\
&-2\nabla_{\gamma}h(X_i,Z_i;\gamma)\nabla_{\gamma}h(X_i,Z_i;\gamma)'h(X_i,Z_i;\gamma)^2\\
&+\nabla_{\gamma}h(X_i,Z_i;\gamma)\nabla_{\gamma}h(X_i,Z_i;\gamma)'\left((D_i - Z_i'\pi_{1}  - X_i'\pi_{2})^2-h(X_i,Z_i;\gamma)^2\right).
\end{align*}
Because $E[(D - Z'\pi_{1}  - X'\pi_{2})^2|X,Z]=h(X,Z_;\gamma)^2$, $E[\Sigma_\gamma]$ simplifies to 
\begin{align*}
E[\Sigma_\gamma]=-2E[\nabla_{\gamma}h(X_i,Z_i;\gamma)\nabla_{\gamma}h(X_i,Z_i;\gamma)'h(X_i,Z_i;\gamma)^2].
\end{align*}
Thus, we have that
\begin{align*}
E\left[\frac{\partial m_i(\phi)}{\partial \phi'}\right]=\begin{bmatrix}
   -E[Z_iZ_i'] & -E[Z_iX_i'] & 0\\
   -E[X_iZ_i'] & -E[X_iX_i'] & 0\\
   0        & 0        & -2E[\nabla_{\gamma}h_i\nabla_{\gamma}h_i'h_i^2]\\
\end{bmatrix}.
\end{align*}
Therefore, a LLN holds for $\frac{\partial m_i(\phi)}{\partial \phi'}$ and $E[\frac{\partial m_i(\phi)}{\partial \phi'}]$ is invertible
because of Assumptions \ref{assumption:rel} and \ref{assumption:phi_estimator}. Therefore, we have
\begin{align*}
0 
&= \frac{1}{n}\sum_{i=1}^n m_i(\phi) 
   + E\left[ 
       \frac{\partial m_i(\phi)}{\partial \phi'} 
     \right]
     (\widehat\phi - \phi) + o_p(\|\widehat\phi - \phi\|)\\
   &= \frac{1}{n}\sum_{i=1}^n m_i(\phi) 
   - \Sigma_{\phi}
     (\widehat\phi - \phi) + o_p(\|\widehat\phi - \phi\|)  ,
\end{align*}
where $\Sigma_{\phi}:=-E\left[ 
       \frac{\partial m_i(\phi)}{\partial \phi'} 
     \right]$. Multiplying through by $\sqrt{n}$ and solving for $\widehat\phi - \phi$ yields
\begin{align*}
\sqrt n(\widehat \phi-\phi) &= \Sigma_{\phi}^{-1} \frac{1}{\sqrt n}\sum_{i=1}^nm_i(\phi) + o_p(1).
\end{align*}
Since the $\Sigma_{\phi}$ matrix above is block diagonal, then we have
\begin{align*}
\Sigma_{\phi}^{-1}= \begin{bmatrix}\begin{bmatrix}
   E[Z_iZ_i'] & E[Z_iX_i'] \\
   E[X_iZ_i'] & E[X_iX_i'] 
   \end{bmatrix}^{-1} & 0\\
     0        & \frac{1}{2}E[\nabla_{\gamma}h(X_i,Z_i;\gamma)\nabla_{\gamma}h(X_i,Z_i;\gamma)'h(X_i,Z_i;\gamma)^2]^{-1}
\end{bmatrix}.
\end{align*}
Then, $\psi_{\phi,i}=\Sigma_{\phi}^{-1}m_i$, and $\sqrt n(\widehat \phi-\phi) =\frac{1}{\sqrt n}\sum_{i=1}^n\psi_{\phi,i} + o_p(1).$ Therefore, $\sqrt n(\widehat \phi-\phi)\overset{d}{\to}\mathcal{N}(0,E[\psi_{\phi}\psi_{\phi}']).$ Thus, $\Omega_\phi = E[\psi_{\phi}\psi_{\phi}'].$

\qed

\subsection{Proof of Theorem \ref{lemma:limiting_distibution}}
The goal is to establish the limiting distribution of the OLS estimators of the following regression given in \eqref{eq:x2_tilde} and \eqref{eq:y_fwl}:
\begin{align} \label{eq:reg:w_d}
  Y & = D\alpha_1 + X'\alpha_2 +  \left( 1 + \sum_{s=1}^{k_{g}} (\theta_{s}^{d}D^{s}+X'^{s}\theta_{s}^{x}) \right)  \left( \sum_{j=1}^{k_v} \mu_jV^j \right)  + U\\
    & = D\alpha_1 + W'\alpha_w + U, \notag
\end{align}
where $U:= Y- E[Y|D, W] $ and is implemented with $V$ replaced by $\widehat V$. We follow Appendix 6A in \cite{Wooldridge10} closely. Recall that $V$ is given by
\begin{align*}
V(\phi) &= \frac{D - Z'\pi_1 - X'\pi_2}{h(X,Z;\gamma)},
\end{align*}
where $\phi:=(\pi_1', \pi_2', \gamma')'$. The estimated counterpart, detailed in Section \ref{section:first_stage}, is
\begin{align*}
    \widehat{V} &= V(\widehat \phi)\\
                &= \frac{D - Z'\widehat \pi_1 - X'\widehat \pi_2}{h(X,Z;\widehat \gamma)}.
\end{align*}

Let $W(\phi)$ be all the regressors in \eqref{eq:reg:w_d} (except $D$), and let $W(\widehat{\phi})=\widehat{W}$ denote those regressors where $V$ is replaced by the generated $\widehat V.$ In terms of the regression in equation \eqref{eq:reg:w_d}, this is 
\begin{align*}
W'\alpha_w=W(\phi)'\alpha_w =  X'\alpha_2+\left( 1 + \sum_{s=1}^{k_{g}} (\theta_{s}^{d}D^{s}+X'^{s}\theta_{s}^{x}) \right)  \left( \sum_{j=1}^{k_v} \mu_jV(\phi)^j \right),
\end{align*}
and
\begin{align*}
\widehat W'\alpha_w =W(\widehat{\phi})'\alpha_w = X'\alpha_2+ \left( 1 + \sum_{s=1}^{k_{g}} (\theta_{s}^{d}D^{s}+X'^{s}\theta_{s}^{x}) \right)  \left( \sum_{j=1}^{k_v} \mu_jV(\widehat{\phi})^j \right).
\end{align*}

Furthermore, let $R_i(\phi):=(D_i,W_i(\phi)')'$ and $R_i(\widehat \phi)=\widehat R_i:=(D_i,\widehat W_i')'$. Also denote $ \alpha := ( \alpha_1, \alpha_w')'$, and $\widehat \alpha := (\widehat \alpha_1, \widehat \alpha_w')'$, where $\widehat \alpha_1$, and $\widehat \alpha_w$ are the OLS estimators of the regression of $Y$ on $D$ and $\widehat W$.

If $\phi$ were known, then the influence function of $\hat \alpha$ would be given by $E[R(\phi)R(\phi)']^{-1}R_i(\phi)U_i$. To account for the influence of estimating $\phi$ we need to evaluate the effect it has on $R(\phi)U=R(\phi)(Y-R(\phi)'\alpha)$. 

For this we need first to compute $J_{\phi,i}=\nabla_\phi R_i(\phi)$, that is, the derivative of $R(\phi)$ with respect to $\phi$, a $dim(R)\times dim(\phi)$ matrix. 

Write \eqref{eq:mom_cond_pi_gamma} succinctly as
\begin{align*}
\frac{1}{n}\sum_{i=1}^nm_i(\widehat \phi )=o_p(n^{-1/2}).
\end{align*}
Combining this with the first order condition for the OLS estimation of $\alpha$, we get
\begin{align*}\frac{1}{n}\sum_{i=1}^n\begin{bmatrix}
m_i(\widehat \phi )\\
R_i(\widehat\phi)(Y_i-R_i(\widehat\phi)'\widehat\alpha)
\end{bmatrix}
=o_p(n^{-1/2}).
\end{align*}
Focusing on the second step estimation, we get using Assumption \ref{assumption:alpha_1_estimator}:
\begin{align*}
0&=\frac{1}{n}\sum_{i=1}^n
R_i(\widehat\phi)(Y_i-R_i(\widehat\phi)'\widehat\alpha)\\
&=\frac{1}{n}\sum_{i=1}^n
R_i(\phi)(Y_i-R_i(\phi)'\widehat\alpha) \\
& \ \ + \frac{1}{n}\sum_{i=1}^n(U_iJ_{\phi,i}- R_i(\phi)\alpha'J_{\phi,i}) (\widehat\phi -\phi ) + o_p(n^{-1/2}),\\
&=\frac{1}{n}\sum_{i=1}^n
R_i(\phi)(Y_i-R_i(\phi)'\widehat\alpha) + E[UJ_\phi- R(\phi)\alpha'J_\phi] (\widehat\phi -\phi ) + o_p(n^{-1/2})\\
&=\frac{1}{n}\sum_{i=1}^n
R_i(\phi)U_i 
- \left(\frac{1}{n}\sum_{i=1}^n R_i(\phi)R_i(\phi)'\right)(\widehat\alpha-\alpha) \\
&\ \ + E\left[UJ_\phi- R(\phi)\alpha'J_\phi\right] (\widehat\phi -\phi ) + o_p(n^{-1/2}).
\end{align*}
Let 
\begin{align*}
\psi_{\alpha,i}=E[RR']^{-1}R_i U_i + E[RR']^{-1}E[UJ_\phi- R\alpha'J_\phi]\psi_{\phi,i}
\end{align*}
and $ \Omega_\alpha:=E[\psi_{\alpha} \psi_{\alpha}']$. Therefore,
\begin{align*} 
\sqrt n (\widehat \alpha-\alpha) &=\frac{1}{\sqrt n}\sum_{i=1}^n\psi_{\alpha,i} + o_p(1),
\end{align*}
and
\begin{align*} 
\sqrt n (\widehat \alpha-\alpha) \stackrel{d}{\rightarrow} \mathcal{N}(0,\Omega_\alpha)
\end{align*}
because by Assumption \ref{assumption:alpha_1_estimator}.$(iii)$ the variance is finite.
\qed

\subsection{Proof of Lemma \ref{lemma_variance}}
The sample counterpart of $\Omega_{\alpha}$ is
\begin{align*}   
\widehat  \Omega_{\alpha} &= \frac{1}{n}\sum_{i=1}^n\hat\psi_{\alpha,i}\hat\psi_{\alpha,i}'.
\end{align*}
with
\begin{align*}
\hat \psi_{\alpha,i}
&=
\left(\frac{1}{n}\sum_{j=1}^n\widehat R_j \widehat R_j'\right)^{-1}\widehat R_i \widehat U_i 
+ 
\left(\frac{1}{n}\sum_{j=1}^n\widehat R_j \widehat R_j'\right)^{-1}
\left(\frac{1}{n}\sum_{j=1}^n\left(\widehat U_j J_{\hat\phi,j}- \widehat R_j\widehat\alpha'J_{\hat\phi,j}\right)\right)\hat \psi_{\phi,i}.
\end{align*}
Here
\begin{align*} 
\widehat U_i &= Y_i - D_i\widehat\alpha_1 - W_i(\widehat\phi)'\widehat\alpha_w = Y_i-\widehat R_i'\widehat \alpha,\\
\hat\psi_{\phi,i} &= \hat\Sigma_\phi^{-1}\hat m_i, \\
\widehat{\Sigma}_{\phi}&=\frac{1}{n}\sum_{i=1}^n\begin{bmatrix}
   Z_iZ_i' & Z_iX_i' & 0\\
   X_iZ_i' & X_iX_i' & 0\\
   0        & 0        & 2[\nabla_{\gamma}h(X_i,Z_i;\widehat\gamma)\nabla_{\gamma}h(X_i,Z_i;\widehat\gamma)'h(X_i,Z_i;\widehat\gamma)^2]\\
\end{bmatrix},\\
\widehat m_i&=\begin{bmatrix}
    Z_i\left(D_i - Z_i'\widehat\pi_1 - X_i'\widehat\pi_2\right)\\
    X_i\left(D_i - Z_i'\widehat\pi_1 - X_i'\widehat\pi_2\right)\\
   \nabla_{\gamma}h(X_i,Z_i;\widehat\gamma) \left(\hat V_i^2-1\right)h(X_i,Z_i;\widehat\gamma)^3
\end{bmatrix}.
\end{align*}
Define
\begin{align*}
\widehat Q_R 
&:= 
\left(\frac{1}{n}\sum_{j=1}^n \widehat R_j \widehat R_j'\right)^{-1}, \\
\widehat b
&:=
\frac{1}{n}\sum_{j=1}^n \left(\widehat U_j J_{\hat\phi,j}- \widehat R_j \widehat\alpha' J_{\hat\phi,j}\right),
\end{align*}
so that 
\begin{align*}
\hat \psi_{\alpha,i}
&=
\widehat Q_R \widehat R_i \widehat U_i 
+ 
\widehat Q_R \widehat b \,\hat \psi_{\phi,i}.
\end{align*}
Therefore, we have that the outer product is 
\begin{align*}
\hat \psi_{\alpha,i}\hat \psi_{\alpha,i}'
&=
\widehat Q_R \widehat R_i \widehat U_i^2 \widehat R_i' \widehat Q_R'
+
\widehat Q_R \widehat R_i \widehat U_i \hat \psi_{\phi,i}'\,\widehat b' \widehat Q_R' \\
&\quad+
\widehat Q_R \widehat b \,\hat \psi_{\phi,i}\,\widehat U_i \widehat R_i' \widehat Q_R'
+
\widehat Q_R \widehat b \,\hat \psi_{\phi,i}\hat \psi_{\phi,i}'\,\widehat b' \widehat Q_R'.
\end{align*}

We start with 
\begin{align*} 
\frac{1}{n}\sum_{j=1}^n\widehat R_j \widehat R_j' = \overset{p}{\to}E[RR'],
\end{align*}
by Assumption \ref{assumption:alpha_1_estimator} $(ii)$, and is not singular by Assumption \ref{assumption:alpha_1_estimator} $(i)$. Similarly, by Assumption \ref{assumption:alpha_1_estimator} $(ii)$ we have
\begin{align*}
\frac{1}{n}\sum_{j=1}^n\left(\widehat U_j J_{\hat\phi,j}- \widehat R_j\widehat\alpha'J_{\hat\phi,j}\right)\overset{p}{\to}E\left[UJ_\phi- R(\phi)\alpha'J_\phi\right].
\end{align*}
Now we show that 
\begin{align*} 
\frac{1}{n}\sum_{i=1}^n\widehat R_i \widehat R_i'\widehat U_i^2 \overset{p}{\to} E[RR'U^2].
\end{align*}
We have
\begin{align*} 
\frac{1}{n}\sum_{i=1}^n\widehat R_i \widehat R_i'\widehat U_i^2 &= \frac{1}{n}\sum_{i=1}^n \widehat R_i \widehat R_i'(Y_i-\widehat R_i'\widehat \alpha)^2\\
&=\frac{1}{n}\sum_{i=1}^n  R_i  R_i'(Y_i - R_i'\widehat \alpha)^2 + \frac{1}{n}\sum_{i=1}^n (\widehat R_i \widehat R_i'-R_i  R_i')(Y_i-\widehat R_i'\widehat \alpha)^2\\
&+ \frac{1}{n}\sum_{i=1}^n R_i  R_i'\left[(Y_i-\widehat R_i'\widehat \alpha)^2- (Y_i- R_i'\widehat \alpha)^2\right]= T_1 + T_2 + T_3.
\end{align*}
We will show that $T_1 \overset{p}{\to}E[R  R'U^2]$, and that $T_2$ and $T_3$ are $o_p(1)$. Consider $T_1$, the first term.
\begin{align*} 
T_1 &= \frac{1}{n}\sum_{i=1}^n  R_i  R_i'(Y_i - R_i'\widehat \alpha)^2\\ &=\frac{1}{n}\sum_{i=1}^n  R_i  R_i'(Y_i - R_i' \alpha - R_i'(\widehat \alpha-\alpha))^2\\
&=\frac{1}{n}\sum_{i=1}^n  R_i  R_i'(Y_i - R_i'\alpha)^2 - \frac{2}{n}\sum_{i=1}^n  R_i  R_i'(Y_i - R_i'\alpha)R_i'(\widehat \alpha-\alpha)\\
&+\frac{1}{n}\sum_{i=1}^n  R_i  R_i'(R_i'(\widehat \alpha-\alpha))^2\\
&=\frac{1}{n}\sum_{i=1}^n  R_i  R_i'(Y_i - R_i'\alpha)^2 + o_p(1)\overset{p}{\to}E[R  R'U^2],
\end{align*}
by Assumption \ref{assumption:alpha_1_estimator} $(ii)$, $\widehat \alpha-\alpha=o_p(1)$, by Assumption \ref{assumption:alpha_1_estimator} $(iii)$, and $E[\|R \|^4]<\infty$.

Now consider the second term, by the Cauchy--Schwarz inequality we have
\begin{align*} 
T_2&=\frac{1}{n}\sum_{i=1}^n (\widehat R_i \widehat R_i'-R_i  R_i')(Y_i-\widehat R_i'\widehat \alpha)^2\\&\leq \left (\frac{1}{n}\sum_{i=1}^n \|\widehat R_i \widehat R_i'-R_i  R_i'\|^2 \right)^{1/2}\left (\frac{1}{n}\sum_{i=1}^n(Y_i-\widehat R_i'\widehat \alpha)^4 \right)^{1/2}\\
&= T_{21}^{1/2} T_{22}^{1/2}.
\end{align*}
We will show that $T_{21}=o_p(1)$ and $T_{22}=O_p(1)$. 

Let $e_i:=\widehat R_i-R_i$. Note that $\widehat R_i\widehat R_i'-R_iR_i'
= R_ie_i' + e_iR_i' + e_ie_i'.$ Hence, using the operator (matrix) norm, $\|ab'\|_{\mathrm{op}}=\|a\|\,\|b\|$, and the triangle inequality,
\begin{equation}\label{eq:DeltaBound}
\|\widehat R_i\widehat R_i'-R_iR_i'\|
\le 2\|R_i\|\,\|e_i\|+\|e_i\|^2,
\qquad
\|\widehat R_i\widehat R_i'-R_iR_i'\|^2
\le C\big(\|R_i\|^2\|e_i\|^2+\|e_i\|^4\big)
\end{equation}
for some constant $C>0$.

To show $T_{21}=o_p(1)$, we use \eqref{eq:DeltaBound}:
\begin{align*} 
T_{21}
=\frac{1}{n}\sum_{i=1}^n \|\widehat R_i\widehat R_i'-R_iR_i'\|^2
\le
C\left\{\Big(\frac{1}{n}\sum_{i=1}^n \|R_i\|^4\Big)^{1/2}
\Big(\frac{1}{n}\sum_{i=1}^n \|e_i\|^4\Big)^{1/2}
+\frac{1}{n}\sum_{i=1}^n \|e_i\|^4\right\}.
\end{align*} 
Therefore, since we have $E\|R_i\|^4<\infty$ and if $\frac{1}{n}\sum_{i=1}^n \|e_i\|^4=o_p(1)$, then $T_{21}=o_p(1)$. Now, to analyze $\frac{1}{n}\sum_{i=1}^n \|e_i\|^4$, we have by an element-by-element application of the mean value theorem $e_i:=R_i(\widehat\phi)-R_i(\phi)=\nabla_\phi R_i(\tilde\phi)(\widehat\phi-\phi)$,
so that $\|e_i\|\leq \|\nabla_\phi R_i(\tilde\phi)\| \|\widehat\phi-\phi\| $ and
\begin{align}\label{eq_r_hat}
\frac{1}{n}\sum_{i=1}^n \|e_i\|^4&\leq \|\widehat\phi-\phi\| ^4\frac{1}{n}\sum_{i=1}^n \|\nabla_\phi R_i(\tilde\phi)\|^4\notag\\
&=O_p(n^{-2})(E[\|J_{\phi}\|^4]+o_p(1)).
\end{align} 
where we assumed that the Jacobian $J_{\phi}$ has finite fourth moments. 

Now we will show that $T_{22}:=\frac{1}{n}\sum_{i=1}^n(Y_i-\widehat R_i'\widehat \alpha)^4=O_p(1)$. We can write
\begin{align*} 
Y_i-\widehat R_i'\widehat \alpha &= U_i - e_i'\widehat\alpha - R_i'(\widehat\alpha -\alpha),
\end{align*}
and for some constant $C$,
\begin{align*} 
(Y_i-\widehat R_i'\widehat \alpha)^4 &\leq C\left\{U_i^4 + (e_i'\widehat\alpha)^4 + (R_i'(\widehat\alpha -\alpha))^4\right\}.
\end{align*}
Averaging gives
\begin{align*} 
T_{22}
=\frac{1}{n}\sum_{i=1}^n(Y_i-\widehat R_i'\widehat\alpha)^4
\le C\left\{
\frac{1}{n}\sum_{i=1}^n(Y_i- R_i' \alpha)^4
+
\frac{1}{n}\sum_{i=1}^n(e_i'\widehat\alpha)^4
+
\frac{1}{n}\sum_{i=1}^n(R_i'(\widehat\alpha -\alpha))^4
\right\}.
\end{align*}

The first term, since $E[U^4]<\infty$, is $O_p(1)$. For the second term, by Cauchy--Schwarz,
\begin{align*} 
(e_i'\widehat\alpha)^4 \le \|\widehat\alpha\|^4\,\|e_i\|^4
\quad\Rightarrow\quad
\frac{1}{n}\sum_{i=1}^n(e_i'\widehat\alpha)^4
\le \|\widehat\alpha\|^4\cdot \frac{1}{n}\sum_{i=1}^n\|e_i\|^4.
\end{align*}
Since $\widehat\alpha=O_p(1)$ and as shown in \eqref{eq_r_hat}, $\frac{1}{n}\sum_{i=1}^n\|e_i\|^4=o_p(1)$, then the second term is $o_p(1)$. For the third term, again by Cauchy--Schwarz,
\begin{align*}
(R_i'(\widehat\alpha -\alpha))^4 \le \|R_i\|^4\,\|\widehat\alpha -\alpha\|^4
\quad\Rightarrow\quad
\frac{1}{n}\sum_{i=1}^n(R_i'(\widehat\alpha -\alpha))^4
\le \|\widehat\alpha -\alpha\|^4\cdot \frac{1}{n}\sum_{i=1}^n\|R_i\|^4.
\end{align*}
Since  $E\|R_i\|^4<\infty$ and $\|\widehat\alpha -\alpha\|^4=O_p(n^{-2})$, then the third term is $o_p(1)$. Therefore, $T_2=o_p(1).$

Finally, consider the third term, by the Cauchy--Schwarz inequality we have
\begin{align*} 
T_{3}&= \frac{1}{n}\sum_{i=1}^n R_i  R_i'\left[(Y_i-\widehat R_i'\widehat \alpha)^2- (Y_i- R_i'\widehat \alpha)^2\right]\\
&\leq \left (\frac{1}{n}\sum_{i=1}^n \|R_i\|^2 \right)^{1/2}\left (\frac{1}{n}\sum_{i=1}^n\left[(Y_i-\widehat R_i'\widehat \alpha)^2- (Y_i- R_i'\widehat \alpha)^2\right]^2 \right)^{1/2}.
\end{align*}
Since the fourth moments of $\|R_i\|$ are finite, then $\frac{1}{n}\sum_{i=1}^n \|R_i\|^2=O_p(1)$. Now, for the next term, consider
\begin{align*} 
(Y_i-\widehat R_i'\widehat \alpha)^2- (Y_i- R_i'\widehat \alpha)^2 &= (Y_i- R_i'\widehat \alpha - e_i'\widehat \alpha)^2- (Y_i- R_i'\widehat \alpha)^2\\
&= (e_i'\widehat \alpha)^2 - 2e_i'\widehat \alpha(Y_i- R_i'\widehat \alpha)\\
&=(e_i'\widehat \alpha)^2 - 2e_i'\widehat \alpha(U_i + R_i'(\widehat \alpha-\alpha)).
\end{align*}
Therefore, for a constant $C$, we get
\begin{align*} 
\left[(Y_i-\widehat R_i'\widehat \alpha)^2- (Y_i- R_i'\widehat \alpha)^2\right]^2 
&\leq C\left\{\|e_i\|^4\|\widehat \alpha\|^4 + \|e_i\|^2\|\widehat\alpha\|^2|U_i|^2 + \|e_i\|^2\|\widehat\alpha \|^2\|R_i\|^2\|\widehat \alpha-\alpha\|^2\right\}.
\end{align*}
Thus,
\begin{align*} 
&\frac{1}{n}\sum_{i=1}^n\left[(Y_i-\widehat R_i'\widehat \alpha)^2- (Y_i- R_i'\widehat \alpha)^2\right]^2 \leq \\
&C\left\{\frac{1}{n}\sum_{i=1}^n\|e_i\|^4\|\widehat \alpha\|^4 +\frac{1}{n}\sum_{i=1}^n \|e_i\|^2\|\widehat\alpha\|^2|U_i|^2 + \frac{1}{n}\sum_{i=1}^n\|e_i\|^2\|\widehat\alpha \|^2\|R_i\|^2\|\widehat \alpha-\alpha\|^2\right\}.
\end{align*}
As shown in \eqref{eq_r_hat}, first term is $o_p(1)$. The second and third terms are also $o_p(1)$. Therefore, $T_3=o_p(1).$ In conclusion, $\frac{1}{n}\sum_{i=1}^n\widehat R_i \widehat R_i'\widehat U_i^2\overset{p}{\to}E[RR'U^2].$

Now, the goal is to show:
\begin{align*} 
\frac{1}{n}\sum_{i=1}^n\hat\psi_{\phi,i}\hat\psi_{\phi,i}'=\hat\Sigma_\phi^{-1}\frac{1}{n}\sum_{i=1}^n\hat m_i\hat m_i'\hat\Sigma_\phi^{-1}\overset{p}{\to}\Sigma_\phi^{-1}E[mm']\Sigma_\phi^{-1} = \Omega_{\phi}.
\end{align*}

By Assumption \ref{assumption:phi_estimator}, $\hat\Sigma_{\phi}^{-1}\overset{p}{\to}\Sigma_{\phi}^{-1}$.

Now, for the outer product $\hat m_i\hat m_i'$, recall that $\check {V}_i = D_i - Z_i'\widehat{\pi}_{1}  - X_i'\widehat{\pi}_{2}$, and denote $\hat c_i:=\nabla_{\gamma}h(X_i,Z_i;\widehat\gamma) \left(\hat V_i^2-1\right)h(X_i,Z_i;\widehat\gamma)^3$, we have that
\begin{align*} 
\widehat m_i\widehat m_i'
=
\begin{bmatrix}
Z_i Z_i' \check {V}_i^2 
& Z_i X_i' \check {V}_i^2 
& Z_i \check {V}_i \hat c_i'\\[0.4em]
X_i Z_i' \check {V}_i^2 
& X_i X_i' \check {V}_i^2 
& X_i \check {V}_i \hat c_i'\\[0.4em]
\hat c_i \check {V}_i Z_i'
& \hat c_i \check {V}_i X_i'
& \hat c_i \hat c_i'
\end{bmatrix}.
\end{align*}
Consider $Z_i Z_i' \check {V}_i^2$, such that 
\begin{align*} 
\frac{1}{n}\sum_{i=1}^nZ_i Z_i' \check {V}_i^2 &= \frac{1}{n}\sum_{i=1}^nZ_i Z_i' ( D_i - Z_i'\widehat{\pi}_{1}  - X_i'\widehat{\pi}_{2})^2 \\
&=\frac{1}{n}\sum_{i=1}^nZ_i Z_i' ( Z_i'\pi_1 + X_i'\pi_2+ h(X_i,Z_i;\gamma)V_i - Z_i'\widehat{\pi}_{1}  - X_i'\widehat{\pi}_{2})^2\\
&=\frac{1}{n}\sum_{i=1}^nZ_i Z_i' \left( Z_i'(\pi_1-\widehat{\pi}_{1}) + X_i'(\pi_2-\widehat{\pi}_{2})+ h(X_i,Z_i;\gamma)V_i\right)^2\\
&=\frac{1}{n}\sum_{i=1}^nZ_i Z_i' h(X_i,Z_i;\gamma)^2V^2 + o_p(1)\\
&\overset{p}{\to}E[Z Z' h(X,Z;\gamma)^2V^2]
\end{align*}
because $E[\|Z\|^4]<\infty$, $E[\|X\|^4]<\infty$, and $E[|h(X,Z;\gamma)V|^4]<\infty$. Similar results hold for $Z_i X_i' \check {V}_i^2$ and $X_i X_i' \check {V}_i^2$. Now consider $Z_i \check {V}_i \hat c_i'$. We have
\begin{align*} 
\frac{1}{n}\sum_{i=1}^nZ_i \check {V}_i \hat c_i' &= \frac{1}{n}\sum_{i=1}^nZ_i \check {V}_i c_i' + \frac{1}{n}\sum_{i=1}^nZ_i \check {V}_i (\hat c_i-c_i)' \\
&=\frac{1}{n}\sum_{i=1}^nZ_i \left( Z_i'(\pi_1-\widehat{\pi}_{1}) + X_i'(\pi_2-\widehat{\pi}_{2})+ h(X_i,Z_i;\gamma)V_i\right) c_i'\\
&+\frac{1}{n}\sum_{i=1}^nZ_i \check {V}_i (\hat c_i-c_i)'\\
&=\frac{1}{n}\sum_{i=1}^n Z_i  h(X_i,Z_i;\gamma)V_i c_i' + \frac{1}{n}\sum_{i=1}^nZ_i \check {V}_i (\hat c_i-c_i)' + o_p(1)
\end{align*}
because by Assumption \ref{assumption:phi_estimator}, the second moments are finite. Now for the second term we have
\begin{align*} 
\left \|\frac{1}{n}\sum_{i=1}^nZ_i \check {V}_i (\hat c_i-c_i)'\right\|^2 \leq \frac{1}{n}\sum_{i=1}^n\left \|Z_i \check {V}_i\right\|^2 \frac{1}{n}\sum_{i=1}^n\left \|\hat c_i-c_i\right\|^2.
\end{align*}
Write $c_i(\phi)$ for the population quantity and $\hat c_i = c_i(\widehat\phi)$,
$c_i = c_i(\phi)$.
We have that it is differentiable in $\phi$ with derivative
$\nabla_\phi c_i(\phi)$ and $E\big[\|\nabla_\phi c_i(\phi)\|^2\big] < \infty.$ By a mean value expansion, for each $i$ there exists $\tilde\phi_i$ on the line
segment between $\widehat\phi$ and $\phi$ such that
\begin{align*}
\hat c_i - c_i
= \nabla_\phi c_i(\tilde\phi_i)(\widehat\phi - \phi).
\end{align*}
Hence
\begin{align*}
\frac{1}{n}\sum_{i=1}^n \|\hat c_i - c_i\|^2
&= \frac{1}{n}\sum_{i=1}^n
\big\|\nabla_\phi c_i(\tilde\phi_i)(\widehat\phi - \phi)\big\|^2 \\
&\le \|\widehat\phi - \phi\|^2
\cdot \frac{1}{n}\sum_{i=1}^n \big\|\nabla_\phi c_i(\tilde\phi_i)\big\|^2.
\end{align*}
Since $\widehat\phi$ is $\sqrt{n}$-consistent,
and $\nabla_\phi c_i(\tilde\phi_i)$ satisfies a uniform law of large numbers, then
\begin{align*}
\frac{1}{n}\sum_{i=1}^n \|\hat c_i - c_i\|^2
= O_p(n^{-1}) \cdot O_p(1) = o_p(1).
\end{align*}
A similar result holds for $X_i \check {V}_i \hat c_i'$. Finally,
we want to show that
\begin{align*}
\frac{1}{n}\sum_{i=1}^n \hat c_i \hat c_i'
\overset{p}{\to}
E[c_i c_i'].
\end{align*}
Decompose
\begin{align*}
\frac{1}{n}\sum_{i=1}^n \hat c_i \hat c_i' - E[c_i c_i']
&=
\left\{\frac{1}{n}\sum_{i=1}^n \hat c_i \hat c_i' - \frac{1}{n}\sum_{i=1}^n c_i c_i'\right\}
+
\underbrace{\left\{\frac{1}{n}\sum_{i=1}^n c_i c_i' - E[c_i c_i']\right\}}_{\overset{p}{\to} 0}.
\end{align*}
The second term is $o_p(1)$ because $E\|c_i\|^2<\infty$.
For the first term, we have
\begin{align*}
\hat c_i \hat c_i' - c_i c_i'
&= (\hat c_i - c_i)\hat c_i' + c_i(\hat c_i - c_i)'.
\end{align*}
Hence
\begin{align*}
\left\|\frac{1}{n}\sum_{i=1}^n \hat c_i \hat c_i' - \frac{1}{n}\sum_{i=1}^n c_i c_i'\right\|
&\le
\frac{1}{n}\sum_{i=1}^n \|(\hat c_i - c_i)\hat c_i'\|
+
\frac{1}{n}\sum_{i=1}^n \|c_i(\hat c_i - c_i)'\| \\
&\le
\frac{1}{n}\sum_{i=1}^n \|\hat c_i - c_i\|\,\|\hat c_i\|
+
\frac{1}{n}\sum_{i=1}^n \|c_i\|\,\|\hat c_i - c_i\|.
\end{align*}
By the Cauchy--Schwarz inequality applied to each sum separately
\begin{align*}
\frac{1}{n}\sum_{i=1}^n \|\hat c_i - c_i\|\,\|\hat c_i\|
&\le
\left(\frac{1}{n}\sum_{i=1}^n \|\hat c_i - c_i\|^2\right)^{1/2}
\left(\frac{1}{n}\sum_{i=1}^n \|\hat c_i\|^2\right)^{1/2},\\
\frac{1}{n}\sum_{i=1}^n \|c_i\|\,\|\hat c_i - c_i\|
&\le
\left(\frac{1}{n}\sum_{i=1}^n \|c_i\|^2\right)^{1/2}
\left(\frac{1}{n}\sum_{i=1}^n \|\hat c_i - c_i\|^2\right)^{1/2}.
\end{align*}

From the previous argument we have $\frac{1}{n}\sum_{i=1}^n \|\hat c_i - c_i\|^2 = o_p(1)$. Moreover, by the uniform law of large numbers, we have $\frac{1}{n}\sum_{i=1}^n \|c_i\|^2 = O_p(1)$ and $\frac{1}{n}\sum_{i=1}^n \|\hat c_i\|^2 = O_p(1)$.
Therefore,
\begin{align*}
\left\|\frac{1}{n}\sum_{i=1}^n \hat c_i \hat c_i' - \frac{1}{n}\sum_{i=1}^n c_i c_i'\right\|
= o_p(1)\cdot O_p(1) + O_p(1)\cdot o_p(1)
= o_p(1).
\end{align*}
Thus, 
\begin{align*}
\frac{1}{n}\sum_{i=1}^n \hat c_i \hat c_i'
\overset{p}{\to}
E[c_i c_i'].
\end{align*}

Finally, we take care of the cross term in $\hat \psi_{\alpha,i}\hat \psi_{\alpha,i}'$. 
We only need to show that
\begin{align*}
\frac{1}{n}\sum_{i=1}^n\widehat R_i \widehat U_i \hat \psi_{\phi,i}'
\overset{p}{\to}
E[RU\psi_{\phi,i}'].
\end{align*}
Define
\begin{align*}
e_i^R &:= \widehat R_i - R_i,\\
e_i^U &:= \widehat U_i - U_i,\\
e_i^\psi &:= \hat \psi_{\phi,i} - \psi_{\phi,i},
\end{align*}
so that
\begin{align*}
\widehat R_i &= R_i + e_i^R,\\
\widehat U_i &= U_i + e_i^U,\\
\hat \psi_{\phi,i} &= \psi_{\phi,i} + e_i^\psi.
\end{align*}
Then
\begin{align*}
\widehat R_i \widehat U_i \hat \psi_{\phi,i}'
&= (R_i + e_i^R)(U_i + e_i^U)(\psi_{\phi,i} + e_i^\psi)'\\
&= R_i U_i \psi_{\phi,i}'
+ e_i^R U_i \psi_{\phi,i}'
+ R_i e_i^U \psi_{\phi,i}'
+ R_i U_i (e_i^\psi)' \\
&
+ e_i^R e_i^U \psi_{\phi,i}'
+ e_i^R U_i (e_i^\psi)'
+ R_i e_i^U (e_i^\psi)'
+ e_i^R e_i^U (e_i^\psi)'.
\end{align*}
Averaging over $i$,
\begin{align*}
\frac{1}{n}\sum_{i=1}^n\widehat R_i \widehat U_i \hat \psi_{\phi,i}'
&=
\underbrace{\frac{1}{n}\sum_{i=1}^n R_i U_i \psi_{\phi,i}'}_{=:S_0}
+ \sum_{k=1}^7 S_k,
\end{align*}
where
\begin{align*}
S_1 &= \frac{1}{n}\sum_{i=1}^n e_i^R U_i \psi_{\phi,i}', &
S_2 &= \frac{1}{n}\sum_{i=1}^n R_i e_i^U \psi_{\phi,i}',\\
S_3 &= \frac{1}{n}\sum_{i=1}^n R_i U_i (e_i^\psi)', &
S_4 &= \frac{1}{n}\sum_{i=1}^n e_i^R e_i^U \psi_{\phi,i}',\\
S_5 &= \frac{1}{n}\sum_{i=1}^n e_i^R U_i (e_i^\psi)', &
S_6 &= \frac{1}{n}\sum_{i=1}^n R_i e_i^U (e_i^\psi)',\\
S_7 &= \frac{1}{n}\sum_{i=1}^n e_i^R e_i^U (e_i^\psi)'.
\end{align*}
By the law of large numbers and the moment conditions in Assumptions \ref{assumption:phi_estimator} and \ref{assumption:alpha_1_estimator}, we have
\begin{align*}
S_0
= \frac{1}{n}\sum_{i=1}^n R_i U_i \psi_{\phi,i}'
\overset{p}{\to}
E[RU\psi_{\phi,i}'].
\end{align*}
To complete the proof, we need to show that $S_k = o_p(1)$ for $k=1,\dots,7$. First, we take care of $e_i^R$, $e_i^U$, and $e_i^\psi$.

By \eqref{eq_r_hat}, we have $\frac{1}{n}\sum_{i=1}^n\|e_i^R\|^4 = o_p(1)$, so that
\begin{align*}
\frac{1}{n}\sum_{i=1}^n\|e_i^R\|^2
\le
\left(\frac{1}{n}\sum_{i=1}^n\|e_i^R\|^4\right)^{1/2}
= o_p(1).
\end{align*}
By arguments analogous to those used for $T_2$ and $T_3$ above, and using $E[U^4]<\infty$, we also obtain
\begin{align*}
\frac{1}{n}\sum_{i=1}^n |e_i^U|^2 = o_p(1).
\end{align*}

For $\psi_{\phi,i}$, recall $\psi_{\phi,i} = \Sigma_\phi^{-1}m_i$ and $\hat\psi_{\phi,i} = \hat\Sigma_\phi^{-1}\hat m_i$, so
\begin{align*}
e_i^\psi
&= \hat\psi_{\phi,i} - \psi_{\phi,i}\\
&= (\hat\Sigma_\phi^{-1}-\Sigma_\phi^{-1})\hat m_i
+ \Sigma_\phi^{-1}(\hat m_i - m_i).
\end{align*}
Then,
\begin{align*}
\frac{1}{n}\sum_{i=1}^n\|e_i^\psi\|^2
&\le
2\|\hat\Sigma_\phi^{-1}-\Sigma_\phi^{-1}\|^2 \frac{1}{n}\sum_{i=1}^n\|\hat m_i\|^2
+ 2\|\Sigma_\phi^{-1}\|^2\frac{1}{n}\sum_{i=1}^n\|\hat m_i - m_i\|^2.
\end{align*}
By Assumption \ref{assumption:phi_estimator}, we have
\begin{align*}
\hat\Sigma_\phi^{-1}\overset{p}{\to}\Sigma_\phi^{-1},\qquad
\frac{1}{n}\sum_{i=1}^n\|\hat m_i\|^2 = O_p(1),\qquad
\frac{1}{n}\sum_{i=1}^n\|\hat m_i - m_i\|^2 = o_p(1),
\end{align*}
so
\begin{align*}
\frac{1}{n}\sum_{i=1}^n\|e_i^\psi\|^2 = o_p(1).
\end{align*}
Additionally, we have
\begin{align*}
\frac{1}{n}\sum_{i=1}^n\|\psi_{\phi,i}\|^2 &= O_p(1),\\
\frac{1}{n}\sum_{i=1}^n\|\hat\psi_{\phi,i}\|^2 &= O_p(1).
\end{align*}
We now bound each $S_k$ with Cauchy--Schwarz and the operator norm, and rely on finite fourth moments.

\textbf{Term $S_1$.}
\begin{align*}
\|S_1\|
&=
\left\|\frac{1}{n}\sum_{i=1}^n e_i^R U_i \psi_{\phi,i}'\right\|
\le
\frac{1}{n}\sum_{i=1}^n \|e_i^R\|\cdot|U_i|\cdot\|\psi_{\phi,i}\|\\
&\le
\left(\frac{1}{n}\sum_{i=1}^n\|e_i^R\|^2\right)^{1/2}
\left(\frac{1}{n}\sum_{i=1}^n U_i^2\|\psi_{\phi,i}\|^2\right)^{1/2}
= o_p(1)\cdot O_p(1) = o_p(1).
\end{align*}

\textbf{Term $S_2$.}
\begin{align*}
\|S_2\|
&=
\left\|\frac{1}{n}\sum_{i=1}^n R_i e_i^U \psi_{\phi,i}'\right\|
\le
\left(\frac{1}{n}\sum_{i=1}^n\|R_i\|^2\right)^{1/2}
\left(\frac{1}{n}\sum_{i=1}^n |e_i^U|^2\|\psi_{\phi,i}\|^2\right)^{1/2}\\
&= O_p(1)\cdot o_p(1) = o_p(1),
\end{align*}
since $E\|R_i\|^4<\infty$ and $\frac{1}{n}\sum |e_i^U|^2 = o_p(1)$.

\textbf{Term $S_3$.}
\begin{align*}
\|S_3\|
&=
\left\|\frac{1}{n}\sum_{i=1}^n R_i U_i (e_i^\psi)'\right\|
\le
\left(\frac{1}{n}\sum_{i=1}^n\|R_i\|^2 U_i^2\right)^{1/2}
\left(\frac{1}{n}\sum_{i=1}^n\|e_i^\psi\|^2\right)^{1/2}\\
&= O_p(1)\cdot o_p(1) = o_p(1),
\end{align*}
by $E\|R_i\|^4<\infty$, $E[U_i^4]<\infty$ and $\frac{1}{n}\sum\|e_i^\psi\|^2 = o_p(1)$.

\textbf{Term $S_4$.}
\begin{align*}
\|S_4\|
&=
\left\|\frac{1}{n}\sum_{i=1}^n e_i^R e_i^U \psi_{\phi,i}'\right\|
\le
\frac{1}{n}\sum_{i=1}^n \|e_i^R\|\cdot|e_i^U|\cdot\|\psi_{\phi,i}\|\\
&\le
\left(\frac{1}{n}\sum_{i=1}^n\|e_i^R\|^2\right)^{1/2}
\left(\frac{1}{n}\sum_{i=1}^n |e_i^U|^2\|\psi_{\phi,i}\|^2\right)^{1/2}
= o_p(1)\cdot o_p(1) = o_p(1).
\end{align*}

\textbf{Term $S_5$.}
\begin{align*}
\|S_5\|
&=
\left\|\frac{1}{n}\sum_{i=1}^n e_i^R U_i (e_i^\psi)'\right\|
\le
\left(\frac{1}{n}\sum_{i=1}^n\|e_i^R\|^2 U_i^2\right)^{1/2}
\left(\frac{1}{n}\sum_{i=1}^n\|e_i^\psi\|^2\right)^{1/2}\\
&= o_p(1)\cdot o_p(1) = o_p(1),
\end{align*}
using $E[U_i^4]<\infty$ and the $o_p(1)$ bounds on $\frac{1}{n}\sum\|e_i^R\|^2$ and $\frac{1}{n}\sum\|e_i^\psi\|^2$.

\textbf{Term $S_6$.}
\begin{align*}
\|S_6\|
&=
\left\|\frac{1}{n}\sum_{i=1}^n R_i e_i^U (e_i^\psi)'\right\|
\le
\left(\frac{1}{n}\sum_{i=1}^n\|R_i\|^2\right)^{1/2}
\left(\frac{1}{n}\sum_{i=1}^n |e_i^U|^2\|e_i^\psi\|^2\right)^{1/2}\\
&\le
O_p(1)\cdot
\left(\frac{1}{n}\sum_{i=1}^n |e_i^U|^2\right)^{1/2}
\left(\frac{1}{n}\sum_{i=1}^n\|e_i^\psi\|^2\right)^{1/2}
= O_p(1)\cdot o_p(1)\cdot o_p(1)
= o_p(1).
\end{align*}

\textbf{Term $S_7$.}
\begin{align*}
\|S_7\|
&=
\left\|\frac{1}{n}\sum_{i=1}^n e_i^R e_i^U (e_i^\psi)'\right\|
\le
\frac{1}{n}\sum_{i=1}^n \|e_i^R\|\cdot|e_i^U|\cdot\|e_i^\psi\|\\
&\le
\left(\frac{1}{n}\sum_{i=1}^n\|e_i^R\|^2\right)^{1/2}
\left(\frac{1}{n}\sum_{i=1}^n |e_i^U|^2\right)^{1/2}
\left(\frac{1}{n}\sum_{i=1}^n\|e_i^\psi\|^2\right)^{1/2}\\
&= o_p(1)\cdot o_p(1)\cdot o_p(1)
= o_p(1).
\end{align*}

Therefore, $S_k = o_p(1)$ for all $k=1,\dots,7$, and
\begin{align*}
\frac{1}{n}\sum_{i=1}^n\widehat R_i \widehat U_i \hat \psi_{\phi,i}'
&=
\frac{1}{n}\sum_{i=1}^n R_i U_i \psi_{\phi,i}' + o_p(1)
\overset{p}{\to} E[RU\psi_{\phi,i}'].
\end{align*}
This establishes
\begin{align*}
\frac{1}{n}\sum_{i=1}^n\widehat R_i \widehat U_i \hat \psi_{\phi,i}'
\overset{p}{\to}
E[RU\psi_{\phi,i}'],
\end{align*}
and completes the proof.
\qed

\newpage

\bibliographystyle{econometrica}
\bibliography{Endog.bib}

\end{document}